\documentclass[journal, onecolumn, draftcls,12pt]{IEEEtran}
\usepackage{amssymb,amsmath,epsfig,graphicx,theorem}
\usepackage{amsfonts}
\usepackage{bm}
\usepackage{arydshln}
\usepackage[normalem]{ulem}
\usepackage[linesnumbered,ruled,vlined]{algorithm2e}
\newtheorem{Theo}{Theorem}

\newtheorem{Remark}{Remark}

\SetKwInput{KwInput}{Input}                
\SetKwInput{KwOutput}{Output}              
\SetKwInput{Initialization}{Initialization}  

\usepackage{epsfig,rotating,setspace,latexsym,amsmath,epsf,amssymb,bm,color}
\usepackage{cite}

\IEEEoverridecommandlockouts

\title{The Capacity of Symmetric Private Information Retrieval under Arbitrary Collusion and Eavesdropping Patterns}

\author{Jiale Cheng, Nan Liu, Wei~Kang%
\thanks{J. Cheng and N. Liu is with the National Mobile Communications Research Laboratory,
Southeast University, Nanjing, China (email: \{jlcheng,nanliu\}@seu.edu.cn). W. Kang are with the School of Information Science and Engineering,
Southeast University, Nanjing, China (email: wkang@seu.edu.cn). }%
\thanks{This work was partially supported by the National Natural Science Foundation of China
under Grants $61971135$, the Research Fund of National Mobile Communications Research Laboratory, Southeast University (No. 2020A03) and Six talent peaks project in Jiangsu Province. 
}
}

\begin{document}

\maketitle
\begin{abstract}
We study the symmetric private information retrieval (SPIR) problem  under arbitrary collusion and eavesdropping patterns for replicated databases. We find its capacity, which is the same as the capacity of the original SPIR problem with the number of databases $N$ replaced by a number $F^*$. The number $F^*$ is the optimal solution to a linear programming problem that is a function of the joint pattern, which is the union of the collusion and eavesdropping pattern. This is the first result that shows how two arbitrary patterns collectively affect the capacity of the PIR problem. We draw the conclusion that for SPIR problems, the collusion and eavesdropping constraints are interchangeable in terms of capacity. As special cases of our result, the capacity of the SPIR problem under arbitrary collusion patterns and the capacity of the PIR problem under arbitrary eavesdropping patterns are also found. 
\end{abstract}

\section{introduction}
In a computer network based on a client-server model, the behavior of information retrieval is often related to privacy leaks. A malicious server that monitors user queries will retrieve the server's history and infer the user's behavior in conjunction with other servers. Simultaneously, the privacy of the database often needs to be protected as well, i.e., the user cannot get more information than the desired message. For data queries, especially in the case of information retrieval that requires high confidentiality, it is necessary to ensure the privacy of the retrieval process in the sense of information theory. 


The problem of private information retrieval (PIR) was first proposed in \cite{chor1995private}, where the user wants to retrieve {\it{a certain bit}} out of a database of $K$ bits, from $N$ servers each storing a replicated version of the database.  The privacy of the user is protected when the queries of the user do not reveal any information about which bit is of interest to any single database. 
The PIR problem was reformulated in \cite{sun2017capacity}, 
where the user wants to retrieve a {\it{sufficiently large}} message from the database.  The goal of query design considered in \cite{sun2017capacity} is to maximize the download efficiency, which is defined as the ratio of the size of the desired message to the total number of downloaded symbols from the servers.  The maximum download efficiency, termed the  {\it{capacity}}, of the PIR problem was shown to be \cite{sun2017capacity}
\begin{align}
C_{\text{PIR}}=\left(1+\frac{1}{N}+\frac{1}{N^2}+\cdots+\frac{1}{N^{K-1}} \right)^{-1}.  \label{CapacityPIR}
\end{align}

When considering the privacy of the database, in addition to the privacy of the user, the symmetric private information retrieval (SPIR) problem was proposed \cite{2000Protecting}. In SPIR problems, in addition to protecting the user's message index from every single server, the servers must collectively prevent the user from gaining any information about the undesired messages of the database. The capacity of the SPIR problem was found to be \cite{Sun2017SPIR} 

\begin{align}
C_{\text{SPIR}} =\left\{ \begin{array}{ll}
1-\dfrac{1}{N} & \quad {\text{if }\rho \geq \dfrac{1}{N-1}} \\ 
0 & \quad {\text{otherwise}}\label{CapacitySPIR}
\end{array},   \right.
\end{align}
where $\rho$ is the amount of common randomness the servers share relative to the message size.
Noted that both $C_{\text{PIR}}$ and $C_{\text{SPIR}}$ increase with the number of servers $N$, since with the help of more servers, the privacy of the user can be hidden better from any single server. Further note that $C_{\text{SPIR}}$ is independent to the number of messages $K$, and we have $C_{\text{SPIR}}=\lim\limits_{K\to \infty}C_{\text{PIR}}$. 

In the scenario where
some subsets of databases may communicate and collude to learn about the message index that is of interest to the user, the colluding PIR and SPIR problems were proposed and studied in  \cite{sun2018colluding} and \cite{wang2018secure}, respectively. The collusion structure studied was a symmetric one, called $T$-colluding servers, where out of the $N$ servers, any up to $T$ number of servers may collude. 
To preserve the privacy of the user under possible collusion among databases, the number of downloaded symbols needs to be increased. It is shown in \cite{sun2018colluding} and \cite{wang2019adversaries} that under $T$-colluding servers, the capacities of the PIR and SPIR problems still take on the same form as (\ref{CapacityPIR}) and (\ref{CapacitySPIR}), respectively, but with $N$ replaced by $\frac{N}{T}$. In other words, when any $T$ databases may collude, the number of effective databases has decreased from $N$ to $\frac{N}{T}$, where $\frac{N}{T}$ does not need to be an integer. 

In addition to the database privacy considered in $T$-colluding SPIR, where database privacy is protected against the querying user, reference \cite{wang2019adversaries} further considers database privacy protection against an external eavesdropper. More specifically, the external eavesdropper has the ability to overhear the queries to and the answers from any $E$ out of $N$ servers. This problem has been termed T-ESPIR, and its capacity was found in \cite{wang2019adversaries} to take on the same form of (\ref{CapacitySPIR}), with $N$ replaced by $\frac{N}{\max(T, E)}$. 

Many other variants of the PIR problem have been studied since \cite{sun2017capacity}, in addition to the scenarios mentioned above, i.e., that of colluding servers, database privacy against the querying user, and database privacy against external eavesdroppers. Due to limited space, we provide the references here without going into the detailed settings and results of each one \cite{Wang2019MDSTSPIR, zhang2019private, tandon2017capacity, banawan2018capacity2, tajeddine2017robust, bitar2018staircase, fanti2015efficient, wang2018capacity, wang2019mismatched, sun2018multiround, Xu2018Subpacket, wei2019capacity, samy2019on, zhou2020capacity-achieving, chen2020the, chen2020the2, chen2020gcsa, xu2019capacity, wei2018capacity, chee2019generalization, lin2019improved, xu2018building, tian2018shannon, fazeli2015codes, blackburn2019pir, blackburn2017pir2, kumar2018local, blackburn2017pir, lavauzelle2018private, chan2015private, yang2002private, sun2016blind, melchor2008fast, fanti2014multi, shah2014one, banawan2018private2, jia2019cross, kumar2019achieving, vajha2017binary, kadhe2017private, wei2018private, li2018single, schrijver2003combinatorial, banawan2019improved, jia2019x, sun2019breaking, kadhe2019equivalence, kazemi2019private, jia2019asymptotic, woolsey2019optimal, zhou2019capacity, zhu2019new, banawan2019capacity, heidarzadeh2018capacityITW, kazemi2019single, heidarzadeh2019single, woolsey2019new, Hsuan2019Weakly, OurArXiv, macwilliams1977theory, feyling1993punctured, ling2004coding, van2012introduction, wang2018epsilon, wang2017secure, tajeddine2018robust, raviv2018private, kumar2018private, yang2018private, kim2017cache, sun2017optimal, d2018lifting, tajeddine2018private2, tian2018capacity, banawan2018private, wang2017linear, banawan2018noisy, lin2018asymmetry, shariatpanahi2018multi, banawan2018multi, abdul2017private, heidarzadeh2018capacity, chen2017capacity, wei2018cache, wei2018fundamental, lin2018mds, kumar2017private, tajeddine2018private, Penas2019local, Guo2019leak, Fazeli2015coded, Li2019repair,  Holzbaur2019linear, zhang2017general, zhang2019subpacket, tajeddine2019colluding, tajeddine2019byzantine, wang2019symmetric, BAN18c, BAN18b, BAN18, ATT18}.

The majority of existing PIR papers study the case of symmetric servers. Symmetry simplifies the problem setting and often lead to capacity results. In practice, however, heterogeneity among servers is prevalent. For example, servers belonging to the same company are more likely to collude, and servers who are located geographically close are more likely to eavesdrop at the same time. In this paper, we study the SPIR problem under the heterogeneous collusion and eavesdropping pattern. More specifically, a heterogeneous colluding pattern may be represented by its maximal colluding sets \cite{tajeddine2017arbitrary}, \cite{zhang2017private}  as $\mathcal{P}_c=\{\mathcal{T}_1, \mathcal{T}_2, \cdots, \mathcal{T}_{M_c}\}$, where the servers in set $\mathcal{T}_m$, $m \in [1:M_c]$ may collude, and there are $M_c$ such colluding sets. Similarly, a heterogeneous eavesdropping pattern may be represented by its maximal eavesdropping sets as $\mathcal{P}_e=\{\mathcal{E}_1, \mathcal{E}_2, \cdots, \mathcal{E}_{M_e}\}$, where the servers in set $\mathcal{E}_m$, $m \in [1:M_e]$ may be tapped by the passive eavesdropper simultaneously, and there are $M_e$ such eavesdropping sets. 

While this is the first paper to study arbitrary eavesdropping patterns, the problem of arbitrary collusion patterns has been studied before. Tajeddine {\it{et. al}}  in \cite{tajeddine2017arbitrary} first proposed PIR problems under arbitrary collusion patterns and studied it for MDS-coded databases, where the database messages are encoded using an $[N,J]$ MDS code, and the coded bits are stored in the $N$ servers. Several other works followed, including \cite{jia2017disjoint} for replicated databases, \cite[Section VII]{zhang2017private} for MDS-coded databases, and some discussions in \cite[Appendix D]{sun2018private}, for both the replicated and MDS-coded databases scenarios. Though collusion patterns are diverse, and at first glance, the problem requires a case-by-case analysis due to the property of each specific collusion pattern \cite{zhang2017private}, reference \cite{Liu2020ACPIR} found a general formula for the PIR capacity that holds for any collusion pattern $\mathcal{P}_c$. The capacity formula was shown to be of the form (\ref{CapacityPIR}), with $N$ replaced by a number $S^*$, 
where $S^*$ is the optimal value of the following linear programming problem
 \begin{align}
\max_{\mathbf{y}} \quad & \mathbf{1}_{N}^T \mathbf{y} \nonumber\\
 \text{subject to} \quad & \mathbf{B}^T_{\mathcal{P}_c} \mathbf{y} \leq\mathbf{1}_M \nonumber\\
 & \mathbf{y} \geq \mathbf{0}_N, \nonumber
 \end{align}
 where $\mathbf{B}_{\mathcal{P}_c}$ is the incidence matrix, of size $N \times M_c$, of the collusion pattern $\mathcal{P}_c$, i.e., if Server $n$ is in the $m$-th colluding set $\mathcal{T}_m$ in $\mathcal{P}_c$, we let the $(n,m)$-th element of $\mathbf{B}_{\mathcal{P}_c}$ be $1$, otherwise, it is zero. $\mathbf{1}_k$ ($\mathbf{0}_k$) is the column vector of size $k$ whose elements are all one (zero).

In this paper, inspired by the proof techniques of \cite{Liu2020ACPIR}, we extend the capacity results found for T-ESPIR problems in \cite{wang2019adversaries} to arbitrary collusion and eavesdropping patterns. We find its capacity, which is the same as the capacity of the original SPIR problem in (\ref{CapacitySPIR}) with the number of databases $N$ replaced by a number $F^*$. The number $F^*$ is the optimal solution to a linear programming problem that is a function of the joint pattern, which is the union of the collusion and eavesdropping pattern. Hence, we draw the conclusion that for SPIR problems, the collusion and eavesdropping constraints are interchangeable in terms of capacity. The result shows that the collusion and eavesdropping pattern affects the capacity of the SPIR problem only through the number $F^*$. This is the first result that shows how two arbitrary patterns collectively affect the capacity of the PIR/SPIR problem. As a special case of our result, the capacity of the SPIR problem under arbitrary collusion patterns and the capacity of the PIR problem under arbitrary eavesdropping patterns are also found.


\section{system model} \label{secSystemModel}
\begin {figure}[htbp]
\centering
\includegraphics [ width =1 \textwidth ]{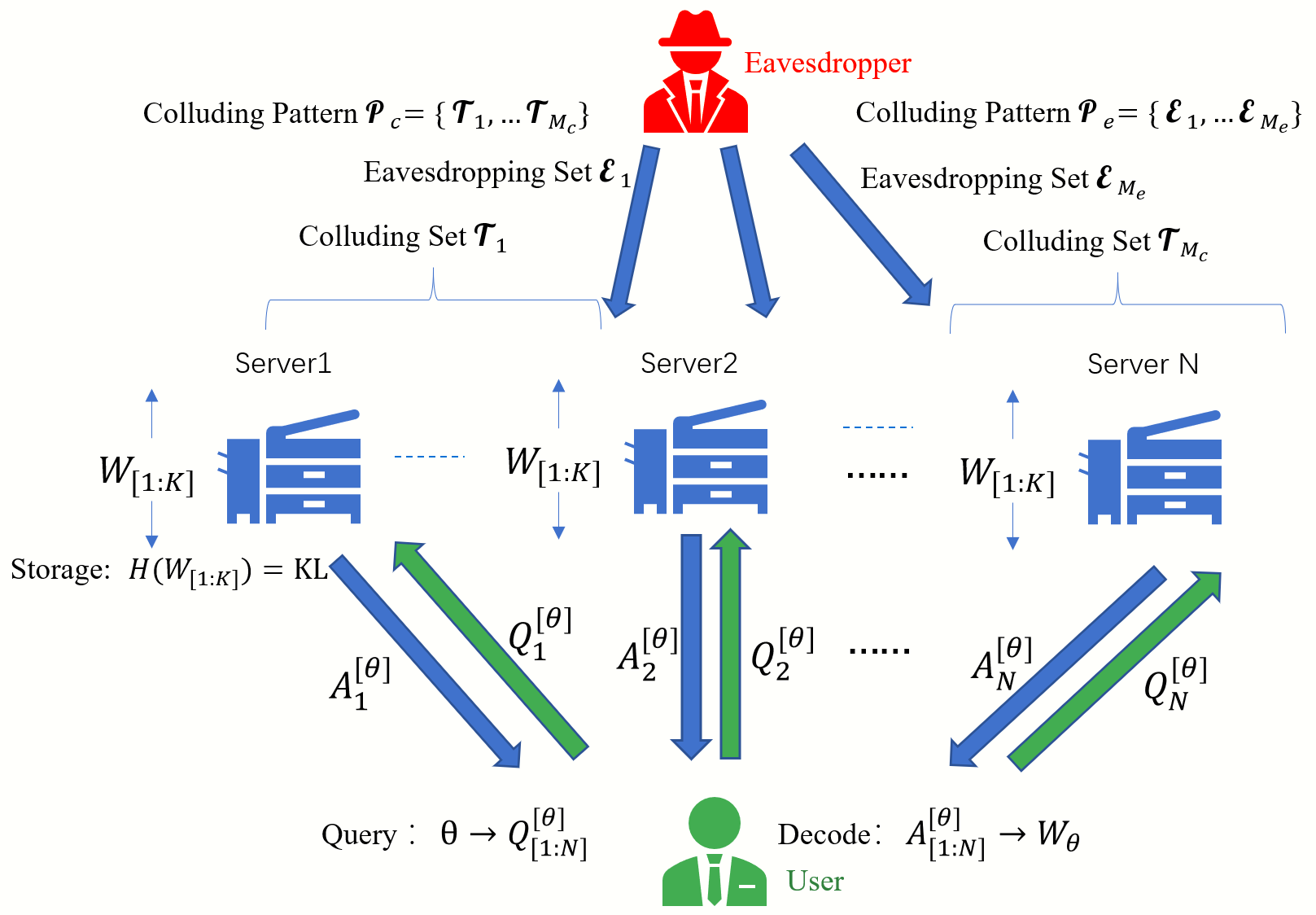}
\caption {System model}
\label{SysMod}
\end {figure}

The system model is as shown in Fig. \ref{SysMod}. Consider the problem where $K$ messages are stored on $N$ replicated databases. The $K$ messages, denoted as $W_k=(W_k^1, W_k^2, \cdots, W_k^L)$, $k \in [1:K]$ are independent and each message consists of $L$ symbols, which are independently and uniformly distributed over a finite field $\mathbb{F}_q$, where $q$ is the size of the field, i.e.,
\begin{align}
H(W_k)&=L, \qquad k=1,...,K, \label{LL}\\
W_{[1:K]}&=H(W_1,...,W_K)=H(W_1)+H(W_2)+\cdots +H(W_K). \nonumber
\end{align}

A user wants to retrieve message $W_\theta$, $\theta \in [1:K]$, by sending designed queries to the databases, where the query sent to the $n$-th database is denoted as $Q_n^{[\theta]}$. Since the queries are designed by the user, who do not know the content of the messages, we have
\begin{align}
I(W_{1:K}; Q_{1:N}^{[\theta]})=0, \quad \forall \theta \in [1:K]. \nonumber
\end{align}
%

In order to protect the privacy of the database from the user and possible eavesdroppers, the servers share a common random variable $S$, which is independent to the messages and queries, and is not known to the users and the eavesdroppers, i.e.:
\begin{equation}
I(S; W_{1:K}, Q_{1:N}^{[\theta]}) = 0. \nonumber 
\end{equation}

Upon receiving the query $Q_{n}^{[\theta]}$, Database $n$ calculates the answer, denoted as $A_{n}^{[\theta]}$, based on the query received $Q_{n}^{[\theta]}$, the messages $W_{1:K}$ and the common randomness $S$, i.e.,
\begin{align}
H(A_{n}^{[\theta]}|Q_{n}^{[\theta]},W_{1:K}, S)=0, \quad \forall n \in [1:N], \theta \in [1:K]. \label{DatabaseHonest}
\end{align}

The queries need to be designed to satisfy the following four conditions:
\begin{enumerate}
\item Correctness of decoding at the user: the user is able to reconstruct the desired message $W_\theta$ 
from all the answers received from the databases, i.e., 
\begin{align}
H(W_\theta|A_{1:N}^{[\theta]},Q_{1:N}^{[\theta]})=0,  \quad \forall \theta \in [1:K]. \label{NoError}
\end{align}

\item Database privacy against the user: the user learns nothing about the undesired messages in the database, i.e.,
\begin{align}
I(W_{\bar{\theta}}; A_{1:N}^{[\theta]},Q_{1:N}^{[\theta]})=0,  \quad \forall \theta \in [1:K]. \label{DatabasePrivacyUser}
\end{align}
Note that in the PIR problem, the above constraint does not need to be satisfied.

\item User privacy under arbitrary server collusion pattern: 
The collusion pattern considered takes on the general form $\mathcal{P}_c=\{\mathcal{T}_1, \mathcal{T}_2, \cdots, \mathcal{T}_{M_c}\}$, where $M_c$ is the number of colluding sets and $\mathcal{T}_m \subseteq [1:N]$, $\forall m \in [1:M_c]$ is the $m$-th colluding set in $\mathcal{P}_c$. The representation $\mathcal{P}_c$ means that the databases in set $\mathcal{T}_m$ may collude, and there are $M_c$ such colluding sets. As an example, for $N=4$ databases, the $2$-colluding case considered by \cite{sun2018colluding} is denoted as $\mathcal{P}_c=\{\{1,2\}, \{1,3\}, \{1,4\}, \{2,3\}, \{2,4\}, \{3,4\}\}$. Note that all databases must appear in at least one element of $\mathcal{P}_c$, because at the very least, the privacy of the user must be preserved at each single database, which is the requirement of the original PIR problem \cite{sun2017capacity}. To protect the privacy of the  user, we require that databases that are in a colluding set can not learn anything about the desired message index $\theta$, i.e., 
\begin{align}
(Q_{\mathcal{T}}^{[1]}, A_\mathcal{T}^{[1]}, W_{1:K}, S) \sim (Q_{\mathcal{T}}^{[\theta]}, A_\mathcal{T}^{[\theta]}, W_{1:K},S), \quad  \forall \theta \in [1:K], \quad \forall  \mathcal{T} \in \mathcal{P}_c.\label{PrivacyConstraint}
\end{align}
We restrict ourselves to $K \geq 2$ in this paper, because in the case of $K=1$, with only 1 message, protecting the user's privacy becomes trivial. We also restrict ourselves to the case where $\mathcal{P}_c$ does not include the set of all servers, as in this case, no scheme can simultaneously achieve protecting both the user's privacy and database privacy against the user \cite{wang2019adversaries}.

\item Database privacy against a possible passive eavesdropper: assume that there is a passive eavesdropper who is interested in the content of the messages, and can eavesdrop on the queries to and the answers from one of the following sets of servers $\mathcal{E}_1,\cdots, \mathcal{E}_{M_e}\subseteq [1: N]$. We denote the eavesdropping pattern as $\mathcal{P}_e=\{\mathcal{E}_1, \mathcal{E}_2, \cdots, \mathcal{E}_{M_e}\}$, where $M_e$ is the number of possible eavesdropping sets and $\mathcal{E}_m \subseteq [1:N]$, $\forall m \in [1:M_e]$ is the $m$-th eavesdropping set in $\mathcal{P}_e$. As an example, for $N=4$ servers, the symmetric eavesdropping of any $3$ servers considered by \cite{wang2019adversaries} is denoted as $\mathcal{P}_e=\{\{1,2,3\}, \{1,3,4\}, \{2,3,4\}, \{1,2,4\}\}$. We require that the eavesdropper knows nothing about the messages in the database, no matter which set in $\mathcal{P}_e$ it taps, i.e.,
\begin{align}
 I(W_{1:K}; A_{ \mathcal {E}}^{[\theta]}, Q_{ \mathcal {E}}^{[\theta]}) = 0, \quad \forall \mathcal {E} \in \mathcal{P}_e.\label{EC}
\end{align}
We restrict ourselves to the case where $\mathcal{P}_e$ does not include the set of all servers, since when the eavesdropper observes the queries to and the answers from all servers, it can definitely decode $W_\theta$, same as the user. Hence, there is no scheme that can achieve database privacy against the passive eavesdropper \cite{wang2019adversaries}. 
\end{enumerate}

For ease of notation, the collusion pattern $\mathcal{P}_c$ and eavesdropping pattern $\mathcal{P}_e$ satisfy the following: 
we only include the maximal set as elements of $\mathcal{P}_c$ $\left(\mathcal{P}_e \right)$. For example, if $\{1,2,3\} \in \mathcal{P}_c$ $\left(\mathcal{P}_e \right)$, then by definition, $\{1,2\}$ is a colluding (eavesdropping) set too. But we do not include $\{1,2\}$ in $\mathcal{P}_c$  $\left(\mathcal{P}_e \right)$ for ease of representation.

Let $\rho$ be the amount of common randomness the servers share relative to the message size, i.e.,
\begin{align}
\rho =\frac{H(S)}{L} \label{Nan05}
\end{align}

For a given common randomness amount $\rho$, the rate of the PIR problem with collusion pattern $\mathcal{P}_c$ and eavesdropping pattern $\mathcal{P}_e$, denoted as $R_\rho(\mathcal{P}_c, \mathcal{P}_e)$, is defined as the ratio between the message size $L$ and the total number of downloaded information from the databases,
i.e.,
\begin{align}
R_\rho(\mathcal{P}_c, \mathcal{P}_e)=  \frac{L}{\sum_{n=1}^{N} H(A_{n}^{[\theta]})} \label{DefineR},
\end{align}
which is not a function of $\theta$ due to the privacy constraint in (\ref{PrivacyConstraint}). 
The capacity of the PIR problem with collusion pattern $\mathcal{P}_c$ and eavesdropping pattern $\mathcal{P}_e$ given common randomness amount $\rho$ is $C_\rho(\mathcal{P}_c, \mathcal{P}_e)=\sup R_\rho(\mathcal{P}_c, \mathcal{P}_e)$, where the supremum is over all possible retrieval schemes.

For the collusion pattern $\mathcal{P}_c$ and eavesdropping pattern $\mathcal{P}_e$, we define a joint pattern $\mathcal{P}_J$ as the maximal set representation of $\mathcal{P}'=\mathcal{P}_c \cup \mathcal{P}_e$. 
For example, for $\mathcal{P}_c=\{\{1,2\},\{3,4\}\}$ and $\mathcal{P}_e=\{\{1,2,3\},\{4\}\}$, the joint pattern $\mathcal{P}_J=\{\{1,2,3\},\{3,4\}\}$. We denote the number of sets in the joint pattern $\mathcal{P}_J$ as $M_J$.

For pattern $\mathcal{P}$ consisting of $M$ sets, we define its corresponding {\it{incidence matrix}} $\mathbf{B}_\mathcal{P}$, of size $N \times M$, as follows: 
%
%
if Server $n$ is in the $m$-th set in $\mathcal{P}$, we let the $(n,m)$-th element of $\mathbf{B}_\mathcal{P}$ be $1$, otherwise, it is zero. For example, pattern $\mathcal{P}=\{\{1,2\},\{2,3\},\{2,4\},\{1,3,4\}\}$ would correspond to an incidence matrix of
\begin{align}
\mathbf{B}_\mathcal{P}=\begin{bmatrix}
1 &0 &0 &1\\
1 & 1 & 1 & 0\\
0 & 1 & 0 & 1\\
0 & 0 & 1 & 1
\end{bmatrix}. \nonumber
\end{align}
The above definition of incidence matrix is applicable to the collusion pattern $\mathcal{P}_c$, the eavesdropping pattern $\mathcal{P}_e$ and their joint pattern $\mathcal{P}_J$. 
Throughout the paper, we will denote the $k \times 1$ column vector of all ones as $\mathbf{1}_k$, and the $k \times 1$ column vector of all zeros as $\mathbf{0}_k$. $\mathbf{I}_k$ is the size $k \times k$ identity matrix and when the size is evident, we write it as $\mathbf{I}$. Similarly, $\mathbf{0}_{k \times i}$ is the size $k \times i$ matrix of all zeros, and when the size is evident, we write it as $\mathbf{0}$. 

\section{Main Results}
The main result on the SPIR capacity under arbitrary collusion and eavesdropping pattern is given in the following theorem. 
\begin{Theo} \label{main}
When $K \geq 2$, the capacity of the SPIR problem under arbitrary collusion pattern $\mathcal{P}_c$ and eavesdropping pattern $\mathcal{P}_e$ is
\begin{align}
C_\rho(\mathcal{P}_c, \mathcal{P}_e) =\left\{ \begin{array}{ll}
1-\dfrac{1}{F^*} & \quad {\text{if }\rho \geq \dfrac{1}{F^*-1}} \\ 
0 & \quad {\text{otherwise}}
\end{array},  \right. \label{CapacityArb}
\end{align}
where $F^*$ is the optimal value of the following linear programming problem (LP1):
\begin{align} 
\text{(LP1)}\quad \max_{\mathbf{y}} \quad & \mathbf{1}_{N}^T \mathbf{y} \nonumber\\
\text{s. t. } \quad & \mathbf{B}_{\mathcal{P}_J}^T \mathbf{y} \leq\mathbf{1}_{M_J} \label{Constraint02}\\
& \mathbf{y} \geq \mathbf{0}_N, \label{Positive01}
\end{align}
where $\mathbf{B}_{\mathcal{P}_J}$ is the incidence matrix, of size $N \times M_J$, of the joint pattern $\mathcal{P}_J$ of collusion pattern $\mathcal{P}_c$ and eavesdropping pattern $\mathcal{P}_e$.
\end{Theo}

Theorem \ref{main} will be proved in the following section. We will first show that $1-\dfrac{1}{F^*}$  is achievable as long as the amount of common randomness relative to the message size satisfies $\rho \geq \dfrac{1}{F^*-1}$. This is achieved by distributing the amount of data queried to each database proportional to the optimal solution $\mathbf{y}^*$ of (LP1). Next, we present a converse theorem that gives an upper bound on capacity as $1-\dfrac{1}{F_2^*}$, where  $F_2^*$ is the optimal value of another linear programming problem (LP2). We further show in the converse that an achievable scheme that protects the user's privacy against the colluding servers in $\mathcal{P}_c$ and protects the database's privacy against the user and the passive eavesdropper in $\mathcal{P}_e$ exists, if and only if the amount of common randomness relative to the message size satisfies $\rho \geq \dfrac{1}{F_2^*-1}$, i.e., if $\rho < \dfrac{1}{F_2^*-1}$, the capacity is zero. Finally, we show that (LP1) and (LP2) are dual problems, which means $F^*=F_2^*$. This concludes the proof that (\ref{CapacityArb}) is the capacity of the SPIR problem under arbitrary collusion patterns $\mathcal{P}_c$ and eavesdropping patterns $\mathcal{P}_e$.

We make a few remarks here regarding the main result.  
\begin{Remark}
\normalfont The result of Theorem \ref{main} has a similar form as the SPIR problem in (\ref{CapacitySPIR}), with $N$ replaced by the $F^*$. So arbitrary eavesdropping and collusion patterns affect the capacity of the SPIR problem by reducing the number of effective servers from $N$ to $F^*$. 
\end{Remark}

\begin{Remark}
\normalfont Theorem \ref{main} shows that the arbitrary collusion and eavesdropping patterns affect the capacity of the SPIR problem only through the optimal solution of linear programming problem (LP1). Hence, the effect of arbitrary collusion and eavesdropping patterns affect the capacity of the SPIR problem through one number $F^*$. 
\end{Remark}

\begin{Remark}
\normalfont Theorem \ref{main} is the first result that shows how two arbitrary patterns collectively affect the capacity of the PIR/SPIR problem. In the SPIR problem of arbitrary collusion and eavesdropping patterns, the capacity $C(\mathcal{P}_c, \mathcal{P}_e)$ is determined by the joint pattern $\mathcal{P}_J$. Hence, it does not rely on the individual collusion pattern $\mathcal{P}_c$ or eavesdropping pattern $\mathcal{P}_e$, as long as their joint pattern $\mathcal{P}_J$ is the same. As a special case, we may switch the two patterns and still get the same capacity.
\end{Remark}

\begin{Remark}
\normalfont
Our result coincides with known capacity results of SPIR problems:
\begin{enumerate}
\item The original SPIR problem \cite{Sun2017SPIR}, with no collusion among servers and no passive eavesdropper, corresponds to the collusion pattern $\mathcal{P}_c=\{\{1\}, \{2\}, \cdots, \{N\}\}$ and eavesdropping pattern $\mathcal{P}_e=\phi$, the empty set. Hence, the joint pattern $\mathcal{P}_J=\mathcal{P}_c$, and its incidence matrix is $\mathbf{B}_{\mathcal{P}_J}=\mathbf{I}_N$. The optimal solution to (LP1) is $\mathbf{y}^*=\mathbf{1}_N$, and the corresponding optimal value is $F^*=N$. Hence, the capacity formula in  (\ref{CapacityArb}) becomes (\ref{CapacitySPIR}), consistent with \cite{Sun2017SPIR}.

\item The T-ESPIR problem \cite{wang2018secure}, where any up to $T$ servers many collude and any up to $E$ servers may be eavesdropped, corresponds to the collusion pattern $\mathcal{P}_c=\{\mathcal{T} \subseteq [1:N]: |\mathcal{T}|=T\}$ and the eavesdropping pattern $\mathcal{P}_e=\{\mathcal{E}\subseteq [1:N]: |\mathcal{E}|=E\}$. Thus, the joint pattern is $\mathcal{P}_J=\{\mathcal{D}\subseteq [1:N]: |\mathcal{D}|=\max\{T,E\}\}$. Hence, the incidence matrix of the joint pattern is consists of $M_J={N\choose \max\{T,E\}}$ columns, each with $\max\{T,E\}$ number of $1$s and $N-\max\{T,E\}$ number of $0$s. It is straightforward to see that the optimal solution to (LP1) is $\mathbf{y}^*=\frac{1}{\max\{T,E\}}\mathbf{1}_N$, and the corresponding optimal value $F^*=\frac{N}{\max\{T,E\}}$. Hence, the capacity formula in (\ref{CapacityArb}) becomes 
\begin{align}
C_{\text{T-ESPIR}} =\left\{ \begin{array}{ll}
 1-\dfrac{\max\{T,E\}}{N} & \quad {\text{if }\rho \geq \dfrac{\max\{T,E\}}{N-\max\{T,E\}}} \\ 
 0 & \quad {\text{otherwise}}
 \end{array},  \right. \nonumber
\end{align}
consistent with \cite{wang2019adversaries}. 
\end{enumerate}
\end{Remark}

\begin{Remark}
\normalfont
The special case of Theorem \ref{main} provides us with the following new PIR capacity results:
\begin{enumerate}
\item The capacity of SPIR under arbitrary collusion pattern $\mathcal{P}_c$: By setting $\mathcal{P}_e=\phi$, we obtain the capacity of SPIR under arbitrary collusion pattern $\mathcal{P}_c$, which is previously unknown,
\begin{align}
C_{\mathcal{P}_c} =\left\{ \begin{array}{ll}
1-\dfrac{1}{S^*} & \quad {\text{if }\rho \geq \dfrac{1}{S^*-1}} \\ 
0, & \quad {\text{otherwise}}
\end{array},  \right. \nonumber
\end{align}
where $S^*$ is the optimal value of the following linear programming problem:
	\begin{align}
	\quad \max_{\mathbf{y}} \quad & \mathbf{1}_{N}^T \mathbf{y} \nonumber\\
	\text{s. t. } \quad & \mathbf{B}_{\mathcal{P}_c}^T \mathbf{y} \leq\mathbf{1}_{M_c} \nonumber\\
	& \mathbf{y} \geq \mathbf{0}_N, \nonumber
	\end{align}
	where $\mathbf{B}_{\mathcal{P}_c}$ is the incidence matrix, of size $N \times M_c$, of the collusion pattern $\mathcal{P}_c$. This is the SPIR version of \cite[Theorem 1]{Liu2020ACPIR}, which characterizes the PIR capacity under arbitrary collusion $\mathcal{P}_c$. Comparing the two results, we see that again, the PIR capacity under arbitrary collusion pattern $\mathcal{P}_c$ converges to the SPIR capacity under the same collusion pattern, as the number of messages $K$ goes to infinity.
	
	\item The capacity of PIR and SPIR under arbitrary eavesdropping pattern $\mathcal{P}_e$: By setting $\mathcal{P}_c=\{\{1\},\cdots, \{N\}\}$, we obtain the capacity of SPIR under arbitrary eavesdropping pattern $\mathcal{P}_e$, which is previously unknown,
	
\begin{align}
C_{\mathcal{P}_e} =\left\{ \begin{array}{ll}
1-\dfrac{1}{Q^*} & \quad {\text{if }\rho \geq \dfrac{1}{Q^*-1}} \\ 
0 & \quad {\text{otherwise}}
\end{array},  \right. \nonumber
\end{align}
where $Q^*$ is the optimal value of the following linear programming problem:
	\begin{align}
	\quad \max_{\mathbf{y}} \quad & \mathbf{1}_{N}^T \mathbf{y} \nonumber\\
	\text{s. t. } \quad & \mathbf{B}_{\mathcal{P}_e}^T \mathbf{y} \leq\mathbf{1}_{M_e} \nonumber\\
	& \mathbf{y} \geq \mathbf{0}_N, \nonumber
	\end{align}
	where $\mathbf{B}_{\mathcal{P}_e}$ is the incidence matrix, of size $N \times M_e$, of the eavesdropping pattern $\mathcal{P}_e$. Interestingly, the corresponding PIR problem is simultaneously solved, as the capacities of the PIR and SPIR problems with arbitrary eavesdropping pattern $\mathcal{P}_e$ are the same, i.e., the protection of database's privacy against the user is free and achieved  without increasing the total download from the servers. The detailed proof of its capacity is in Section \ref{Nan21}.
\end{enumerate}
\end{Remark}


\section{Proofs}
\subsection{Achievability} \label{SecAch}
Recall that for each joint pattern $\mathcal{P}_J$ of collusion pattern $\mathcal{P}_c$ and eavesdropping pattern $\mathcal{P}_e$, there is a corresponding incidence matrix $\mathbf{B}_{\mathcal{P}_J}$, as defined in Section \ref{secSystemModel}. Let $\mathbf{y}=\begin{bmatrix} y_1 & y_2& \cdots &y_N \end{bmatrix}^T$ be a feasible and rational solution of (LP1), i.e., $\mathbf{y}$ consists of rational elements, and it satisfies the constraints (\ref{Constraint02}) and (\ref{Positive01}). Let the value of the objective function in (LP1) corresponding to $\mathbf{y}$ be $F$, i.e., $F=\sum_{n=1}^N y_n$. Then, we have the following achievability theorem. 
 \begin{Theo} \label{TheoremAchAnyY}
Consider the SPIR problem with collusion pattern $\mathcal{P}_c$ and eavesdropping pattern $\mathcal{P}_e$ ,
Suppose $\mathbf{y}$ is a rational and feasible solution of (LP1) and $F=\mathbf{1}_N^T \mathbf{y}>1$. Then the following rate is achievable, i.e., 
\begin{align}
C_\rho(\mathcal{P}_c,\mathcal{P}_e) \geq 1-\frac{1}{F}\label{AchRateAnyY}
\end{align}
when the amount of common randomness $\rho$ satisfies 
\begin{align}
\rho  \geq \frac{1}{F-1}. \label{Nan07}
\end{align}
\end{Theo}
\begin{IEEEproof}
The details of the proof of Theorem \ref{TheoremAchAnyY} is provided in Appendix \ref{AppAch}, along with an illustrating example. The proof follows similarly to \cite[Section V.A]{wang2019adversaries}, and we note the difference here: 1) In place of $N$ in  \cite[Section V.A]{wang2019adversaries}, we have $\bar{L}$, which will be chosen such that the number of symbols downloaded from each of the servers is an integer.  Such an $\bar{L}$ can be found since $\mathbf{y}$ is rational. 2) In place of $\max\{T,E\}$ in \cite[Section V.A]{wang2019adversaries}, we have $\frac{\bar{L}}{F}$. 3) Rather than generating $N$ queries with each server receiving 1 query, the user generates $\bar{L}$ number of queries, and distribute the queries among the databases proportionally according to $(\mathbf{y}, F)$, more specifically, the number of queries to Server $n$ is based on the proportion $\frac{y_n}{F}$, $n \in [1:N]$. 
\end{IEEEproof}
\begin{Remark}
\normalfont
The achievable scheme proposed is based on recognizing that in both colluding and eavesdropping sets, the number of queries/answers the servers or the eavesdropper see is crucial. When each colluding set of servers collectively see too many queries, the index of the user's desired message can be deduced, and therefore, the user's privacy is leaked.  When the eavesdropper sees too many answers, the number of common randomnesses is not enough to protect all the messages, and the database privacy is leaked. Thus, in the design of the achievable scheme, we need to balance the amount of queries each server receives, and correspondingly, the number of answers it sends. This balance is represented by (\ref{Constraint02}), where by considering the joint pattern of collusion and eavesdropping, it is assured that both colluding servers and the passive eavesdropper do not see too many queries/answers. Considering the joint pattern, i.e., (\ref{Constraint02}), may seem like a condition that is too strong, i.e., sufficient but not necessary, but as will be shown in the converse, this is in fact optimal. 
\end{Remark}

\begin{Remark}
\normalfont
 It is easy to see that $\mathbf{y}=\frac{1}{N} \mathbf{1}_N$ is a feasible and rational solution of (LP1). 
 The corresponding cost function $F=\mathbf{1}_N^T \mathbf{y}=1$. Hence, the optimal cost function of (LP1) satisfies $F^* \geq 1$. The equality happens if and only if the collusion pattern or the eavesdropping pattern includes the set of all servers, which is not under consideration in this paper, as explained in Section \ref{secSystemModel}. Hence, for the cases of $\mathcal{P}_c$ and $\mathcal{P}_e$ considered in this paper, we always have $F^*>1$. 
\end{Remark}

Note that the right-hand side (RHS) of (\ref{AchRateAnyY}) is an increasing function of $F$ and the RHS of (\ref{Nan07}) is a decreasing function of $F$. Based on the result of Theorem \ref{TheoremAchAnyY}, to find the largest possible achievable rate using the smallest amount of common randomness, we should find the maximum $F=\sum_{n=1}^N y_n$ achievable over all $\mathbf{y}$ satisfying (\ref{Constraint02}) and (\ref{Positive01}). Applying Theorem \ref{TheoremAchAnyY} for the optimal solution of (LP1), denoted as $(\mathbf{y}^*, S^*)$, and noting that $\mathbf{y}^*$ is rational due to the fact that the objective function and the linear constraints in (LP1) are both with integer coefficients,  the rate of Theorem \ref{main} is achievable.

\subsection{Converse} \label{SecConverse}
Recall that for the joint pattern $\mathcal{P}_J$ of collusion pattern $\mathcal{P}_c$ and eavesdropping pattern $\mathcal{P}_e$, there is a corresponding incidence matrix $\mathbf{B}_{\mathcal{P}_J}$, as defined in Section \ref{secSystemModel}. 
 Consider the following linear programming problem, which will be called (LP2),
 \begin{align}
\text{(LP2)} \qquad \min_{\mathbf{x}} \quad & \mathbf{1}_{M}^T \mathbf{x} \nonumber\\
 \text{subject to} \quad & \mathbf{B}_{\mathcal{P}_J} \mathbf{x} \geq \mathbf{1}_N \label{Constraint01}\\
 & \mathbf{x} \geq \mathbf{0}_{M}. \label{Positive02}
 \end{align}
 
 Let $\mathbf{x}=\begin{bmatrix} x_1 & x_2& \cdots &x_M \end{bmatrix}^T$ be a feasible and rational solution of (LP2), i.e., $\mathbf{x}$ consists of rational elements, and it satisfies the constraints (\ref{Constraint01}) and (\ref{Positive02}). Let the value of the objective function in (LP2) corresponding to $\mathbf{x}$ be $F_2$, i.e., $F_2=\sum_{m=1}^M x_m$.  

 We have the following converse theorem. 
 \begin{Theo} \label{TheoremConverseAnyX}
Consider the SPIR problem with collusion pattern $\mathcal{P}_c$ and eavesdropping pattern $\mathcal{P}_e$. Let $\mathcal{P}_J$ be the joint pattern, whose incidence matrix is $\mathbf{B}_{\mathcal{P}_J}$. Suppose $\mathbf{x}$ is a rational and feasible solution of (LP2) and $F_2=\mathbf{1}_M^T \mathbf{x}$. Then, the capacity of the SPIR problem is upper bounded by 
\begin{align}
C(\mathcal{P}_c, \mathcal{P}_e) \leq 1-\frac{1}{F_2}.\label{GeneralConverse}
\end{align}
Furthermore, for an achievable scheme to exist, the amount of common randomness shared by the servers relative to the message size must satisfy 
\begin{align}
\rho \geq \frac{1}{F_2-1}. \label{Nan14}
\end{align} 
i.e., if (\ref{Nan14}) is not satisfied, then the capacity of the SPIR problem is zero. 
\end{Theo}
\begin{IEEEproof}
The proof follows by combining the converse results of SPIR with $T$-colluding and $E$-eavesdropping \cite{wang2019adversaries} with the converse results of PIR with arbitrary collusion pattern \cite{Liu2020ACPIR}. The details are provided in Appendix \ref{proof01}. 

The reason why only the joint pattern matters is because, for each eavesdropping set $\mathcal{N} \in \mathcal{P}_e$, we have 
\begin{align}
H(A_{\mathcal{N}}^{[k]}|W_k, Q_{\mathcal{N}}^{[k]})=H(A_\mathcal{N}^{[k]}|Q_{\mathcal{N}}^{[k]}). \label{Nan20}
\end{align}

Furthermore, in SPIR problems, for each colluding set, i.e., $\forall \mathcal{N} \in \mathcal{P}_c$, (\ref{Nan20}) is still true \cite{wang2019adversaries}. Hence, in SPIR problems, colluding and eavesdropping sets play the same role in the converse proof. This is why only the joint pattern, i.e., the union of the colluding and eavesdropping pattern matters. 
\end{IEEEproof}

Note that the RHS of (\ref{GeneralConverse}) is an increasing function of $F_2$ and the RHS of (\ref{Nan14}) is a decreasing function of $F_2$. Based on the result of Theorem \ref{TheoremConverseAnyX}, to find the tightest upper bound on capacity and the tightest lower bound on the common randomness, we should find the minimum $F_2=\sum_{m=1}^M x_m$ achievable over all $\mathbf{x}$ satisfying (\ref{Constraint01}) and (\ref{Positive02}). Applying Theorem \ref{TheoremConverseAnyX} for the optimal solution of (LP2), denoted as $(\mathbf{x}^*, S_2^*)$, and noting that $\mathbf{x}^*$ is rational due to the fact that the objective function and linear constraints in (LP2) are both with integer coefficients,  we have 
\begin{align}
C(\mathcal{P}_c, \mathcal{P}_e) \leq 1-\frac{1}{F_2^*}\nonumber
\end{align}
and the amount of common randomness shared by the servers relative to the message size must satisfy 
\begin{align}
\rho \geq \frac{1}{F_2^*-1}. \label{Nan15}
\end{align}

\subsection{Capacity} \label{SecCapacity}
In Section \ref{SecAch}, we have shown that when the amount of common randomness relative to the message size satisfies $\rho \geq \frac{1}{F^*-1}$, the SPIR capacity satisfies $C(\mathcal{P}_c, \mathcal{P}_e) \geq 1-\frac{1}{F^*} $.  In Section \ref{SecConverse}, we have shown that when $\rho \geq \frac{1}{F_2^*-1}$, the SPIR capacity satisfies  $C(\mathcal{P}_c, \mathcal{P}_e) \leq 1-\frac{1}{F_2^*}$, furthermore, when $\rho < \frac{1}{F_2^*-1}$, the SPIR capacity is zero. Recall that
$F^*$  and $F_2^*$ are the optimal solutions to (LP1) and (LP2), respectively.

If $F^*=F_2^*$, then the achievability and converse results meet, yielding the capacity result of Theorem \ref{main}. This is indeed true, as (LP1) and (LP2) are actually dual problems of each other, which means $F^*=F_2^*$.

Hence, we have found the capacity of the SPIR problem under arbitrary collusion pattern $\mathcal{P}_c$ and eavesdropping pattern $\mathcal{P}_e$, as described in Theorem \ref{main}. 

\subsection{PIR problem with arbitrary eavesdropping pattern $\mathcal{P}_e$} \label{Nan21}
Following the proof of our main result, we may get the following theorem on the {\it{PIR}} capacity with arbitrary eavesdropping pattern $\mathcal{P}_e$.
\begin{Theo} \label{sub-result}
When $K \geq 2$, the capacity of the PIR problem under arbitrary eavesdropping pattern $\mathcal{P}_e$ is

\begin{align}
C_\rho^{\text{PIR}} (\mathcal{P}_e) =\left\{ \begin{array}{ll}
1-\dfrac{1}{F^*}, & \quad {\rho \geq \dfrac{1}{F^*-1}} \\ 
0, & \quad {\text{otherwise}}
\end{array},  \right. \nonumber
\end{align}

where $F^*$ is the optimal value of the following linear programming problem (LP1):
\begin{align} 
\text{(LP1)}\quad \max_{\mathbf{y}} \quad & \mathbf{1}_{N}^T \mathbf{y} \nonumber\\
\text{s. t. } \quad & \mathbf{B}_{\mathcal{P}_e}^T \mathbf{y} \leq\mathbf{1}_{M_e} \nonumber\\
& \mathbf{y} \geq \mathbf{0}_N, \nonumber
\end{align}
where $\mathbf{B}_{\mathcal{P}_e}$ is the incidence matrix, of size $N \times M_e$, of the eavesdropping pattern $\mathcal{P}_e$.
\end{Theo}

The above theorem studies the PIR problem, which does not protect the databases' privacy against the user, i.e., (\ref{DatabasePrivacyUser}) does not need to be satisfied. In this case, (\ref{Nan20}) still holds for all the eavesdropping sets, and therefore, the converse holds for the eavesdropping pattern $\mathcal{P}_e$. In terms of achievability, we still employ the same achievability used for SPIR, i.e., even though (\ref{DatabasePrivacyUser}) does not need to be satisfied, the proposed achievability still satisfies (\ref{DatabasePrivacyUser}). Coupled with the converse results, we see that in PIR problems with arbitrary eavesdropping pattern $\mathcal{P}_e$, databases' privacy against the user can be protected for free, i.e., without incurring an increase in the download cost.

PIR, and not SPIR, problems with arbitrary colluding {\it{and}} eavesdropping patterns are still open. There, the colluding pattern and eavesdropping patterns will play different roles, and how they interact with each other is part of ongoing research.

 \section{Conclusions}
 We have found the capacity of the SPIR problem under arbitrary collusion patterns $\mathcal{P}_c$ and eavesdropping pattern $\mathcal{P}_e$. We first link the achievable SPIR rate and its converse to the solutions of two linear programming problems. Then, we show that the two different linear programming problems have the same optimal value. As a result, the achievable SPIR rate and its converse meet, yielding the capacity. From our results, it can be seen that the collusion pattern and eavesdropping pattern do not matter individually, it is their joint pattern that determines the SPIR capacity of the problem.


\appendices

%

\section{Proof of Theorem \ref{TheoremAchAnyY}} \label{AppAch}

\subsection{An Illustrative Example: $K=3, N=5, \mathcal{P}_c=\{\{1,2\},\{1,4\},\{2,4\},\{3,4\},\{5\}\}, \mathcal{P}_e=\{\{1,2,3\},\{2,4\},\{5\}\}$}

The joint pattern is $\mathcal{P}_J=\{\{1,2,3\},\{1,4\},\{2,4\},\{3,4\},\{5\}\}$, and its corresponding incidence  matrix is:
\begin{equation}
 \mathbf {B}_{\mathcal{P}_J} = \begin{bmatrix} 1& 1& 0& 0& 0\\
1& 0 &1& 0 &0 \\
1 &0 &0 &1 &0\\
0 &1 &1 &1 &0\\
0 &0& 0 &0& 1\end{bmatrix}.\nonumber
\end{equation}

A feasible solution to (LP1) is $\mathbf{y}=\begin{bmatrix}\dfrac{1}{3} & \dfrac{1}{3} & \dfrac{1}{3}& \dfrac{2}{3}& 1 \end{bmatrix}^T$, and the corresponding cost is $F=\frac{8}{3}$. The solution $\mathbf{y}$ is in fact the optimal solution of (LP1) but its optimality is not used here. 

Pick $\bar{L}$ such that the following numbers are integers: $\bar{L}$, $\frac{\bar{L}}{F}$, $\frac{\bar{L}}{F} y_n$, $n \in [1:N]$. The smallest $\bar{L}$ that satisfies this is $8$. 

Suppose each message has length $L=\bar{L} \left(1-\frac{1}{F} \right)=5$. We stack the symbols of all $K=3$ messages into a $KL=15$ column vector, i.e.,
\begin{align}
\mathbf{W}=\begin{bmatrix} W_1^1& W_1^2 & \cdots &W_1^5 & \cdots & W_3^1& \cdots & W_3^5 \end{bmatrix}^T \nonumber
\end{align}

The queries are generated at the user in the following way. First, the user generates $\frac{\bar{L}}{F}=3$ many independent and random  column vectors $U_1, U_2, U_{3}$, each with length $KL=15$. The queries generated by the users are given as
\begin{align}
\begin{bmatrix} \bar{Q}_1 & \bar{Q}_2 & \cdots & \bar{Q}_{8}\end{bmatrix} \triangleq \begin{bmatrix} U_1 & U_2  & U_{3} \end{bmatrix} \cdot \mathbf{G}_{\left(8, 3 \right)}+ \begin{bmatrix} \mathbf{0} &\cdots & \mathbf{0} & e_1^{[\theta]} & \cdots & e_{5}^{[\theta]} \end{bmatrix} \label{NanNan01}
\end{align}
where $\mathbf{G}_{\left(8,3\right)}$ is the generating matrix of an $\left(8,3 \right)$-GRS code, $e_k^{[\theta]}$ is a column vector of length $15$ where only the $\left(15(\theta-1)+k\right)$-th element is $1$, and all other elements are $0$. The number of zero vectors in the second term of the RHS of (\ref{NanNan01}) is $\bar{L}-L=3$. The total number of query column vectors in (\ref{NanNan01}) is $\bar{L}=8$, and these will be distributed to the servers according to $(\mathbf{y}, F)$, more specifically, the number of queries to server $n$ is $\frac{y_n}{F} \bar{L}$, i.e., Servers $1$, $2$, $3$ receives $\bar{Q}_1$, $\bar{Q}_2$, $\bar{Q}_3$, respectively, Server $4$ receives $(\bar{Q}_4,\bar{Q}_5)$ and Server 5 receives $(\bar{Q}_6,\bar{Q}_7, \bar{Q}_8)$.

We now show that the queries generated above protects user privacy, i.e., it satisfies (\ref{PrivacyConstraint}). Recall that the colusion pattern is $\mathcal{P}_c=\{\{1,2\},\{1,4\},\{2,4\},\{3,4\},\{5\}\}$. The number of query vectors colluding set $\{1,2\}$ sees is 2, and the number of query vectors colluding sets $\{1,4\},\{2,4\},\{3,4\},\{5\}$ see is 3 each. Thus, we may conclude that the number of query vectors each colluding set sees is no more than 3. 
Due to the MDS property of the $\left(8,3 \right)$-GRS code, any 3 out of $\bar{Q}_1, \cdots, \bar{Q}_8$ are independent and uniformly distributed. Hence, user privacy is preserved against any colluding set of servers in $\mathcal{P}_c=\{\{1,2\},\{1,4\},\{2,4\},\{3,4\},\{5\}\}$.

Next, we describe server answers based on the query vector it receives. The $N=5$ servers share a common randomness vector $\mathbf{S}=\begin{bmatrix} S_1 & S_2  &S_3\end{bmatrix}$. Thus, the common randomness shared relative to the message size is $\rho=\frac{3}{5}$, which is equal to $\frac{1}{F-1}$. Each server calculates the following row vector
$
\bar{\mathbf{S}}=\begin{bmatrix} \bar{S}_1 & \bar{S}_2 & \cdots &\bar{S}_{8}\end{bmatrix} \triangleq \mathbf{S} \cdot \mathbf{G}_{\left(8,3 \right)}
$. Server $n$ calculates the inner product of each query vector it receives with the message vector $\mathbf{W}$, and then adds the corresponding elements from $\bar{\mathbf{S}}$ and sends it to the user, i.e., let 
\begin{align}
\bar{A}_n \triangleq \bar{Q}_n^T \mathbf{W}+\bar{S}_n, \quad n \in [1:8] \label{Nan06}
\end{align}
Servers $1$, $2$, $3$ sends $\bar{A}_1$, $\bar{A}_2$, $\bar{A}_3$, respectively, Server $4$ sends $(\bar{A}_4, \bar{A}_5)$, and Server $5$ sends $(\bar{A}_6, \bar{A}_7, \bar{A}_8)$. 

Now, we verify that the proposed scheme satisfies the decoding constraint at the user, i.e., (\ref{NoError}), the database privacy constraint at the user, i.e., (\ref{DatabasePrivacyUser}) and the database privacy constraint at the passive eavesdropper, i.e., (\ref{EC}). To do this, it is useful to write the server answers in a different form by noticing that both the random vectors $U_1, U_2, U_3$ and the common randomness $S_1, S_2, S_3$ are encoded by the same $\left(8,3 \right)$-GRS code with generating matrix $\mathbf{G}_{(8,3)}$, i.e., the answers sent by the servers may be written in the following matrix form
\begin{align}
&\begin{bmatrix} \bar{A}_1 & \bar{A}_2 & \cdots & \bar{A}_{8}\end{bmatrix} \nonumber\\
= & \begin{bmatrix} \mathbf{W}^T U_1+S_1 & \mathbf{W}^T U_2+S_2 &  \mathbf{W}^T U_3+S_3\end{bmatrix} \cdot \mathbf{G}_{\left(8,3 \right)}+\begin{bmatrix} 0  & \cdots & 0 & W_\theta^1  & \cdots & W_\theta^5 \end{bmatrix}
\nonumber \\
= & \begin{bmatrix} X_1 &X_2 &X_{3} &W_\theta^1  & \cdots & W_\theta^5  \end{bmatrix} \cdot \begin{bmatrix} \mathbf{G}_{\left(8,3\right)}\\ \mathbf{0}_{5 \times 3} \quad \mathbf{I}_5\end{bmatrix}\nonumber
\end{align}
where we have defined $X_l=\mathbf{W}^T U_l+S_l$, $ l \in \left[1:3\right]$. 

The decoding constraint at the user is satisfied, as upon receiving $\bar{A}_l$, $l \in [1:8]$, the user may decode 
$\begin{bmatrix} X_1 &X_2 &X_{3} &W_\theta^1  & \cdots & W_\theta^5  \end{bmatrix}$ 
because 
$\begin{bmatrix} \mathbf{G}_{\left(8,3\right)}\\ \mathbf{0}_{5 \times 3} \quad \mathbf{I}_5\end{bmatrix}$ is invertible due to the MDS property of the generating matrix $\mathbf{G}_{\left(8,3 \right)}$. The database privacy against the user is satisfied because apart from its decoded message $W_\theta$, it can further decode $X_1$, $X_2$, $X_3$, each of which includes a common randomness $S_1$, $S_2$ or $S_3$ to protect the other $K-1=2$ messages.  Finally, we check that the proposed scheme protects database privacy against the passive eavesdropper. Recall that the eavesdropping pattern is $\mathcal{P}_e=\{\{1,2,3\},\{2,4\},\{5\}\}$. Hence, the number of answers each eavesdropping set $\{1,2,3\}$, $\{2,4\}$ and $\{5\}$ sees is equal to 3. From (\ref{Nan06}), we see that each answer $\bar{A}_k$ is protected by a randomness $\bar{S}_k$, and as long as the eavesdropper sees no more than $3$ answers, the corresponding 3 $\bar{S}$s it sees are linearly independent, due to the MDS property of the generating matrix $\mathbf{G}_{\left(8,3 \right)}$. Hence, the  inner product of the corresponding query vector and message vector is fully protected against the passive eavesdropper who observes answers from servers in each eavesdropping set in $\mathcal{P}_e=\{\{1,2,3\},\{2,4\},\{5\}\}$.

\subsection{General Achievability Scheme for arbitrary number of messages $K$, arbitrary number of databases $N$ and arbitrary joint pattern $\mathcal{P}_J$ of collusion pattern $\mathcal{P}_c$ and eavesdropping pattern $\mathcal{P}_e$}

 Let $\mathbf{y}=\begin{bmatrix} y_1 & y_2& \cdots &y_N \end{bmatrix}^T$ be a feasible and rational solution of (LP1), i.e., $\mathbf{y}$ consists of rational elements, and it satisfies the constraints (\ref{Constraint02}) and (\ref{Positive01}). Let the value of the objective function in (LP1) corresponding to $\mathbf{y}$ be $F$, i.e., $F=\sum_{n=1}^N y_n$. Furthermore, $\mathbf{y}$ satisfies $F>1$. 
 
 The encoding of the messages follows the scheme in \cite[Section V.A]{wang2019adversaries} closely with $N$ replaced by $\bar{L}$, and $\max\{T,E\}$ replaced with $\frac{\bar{L}}{F}$. For completeness, we state the scheme here. 

Pick $\bar{L}$ such that the following numbers are integers: $\bar{L}$, $\frac{\bar{L}}{F}$, $\frac{\bar{L}}{F} y_n$, $n \in [1:N]$. 
Note that the above involves $N+2$ numbers which is finite. Such an $\bar{L}$ can be found because $F$ and $y_n$, $n \in [1:N]$ are rational numbers.

Suppose each message has length $L=\bar{L} \left(1-\frac{1}{F} \right)$. Note that $L$ thus defined is positive, because $\mathbf{y}$ satisfies $F>1$.
We stack the symbols of all $K$ messages into a $KL$ column vector, i.e.,
\begin{align}
\mathbf{W}=\begin{bmatrix} W_1^1& W_1^2 & \cdots &W_1^L & \cdots & W_K^1& \cdots & W_K^L \end{bmatrix}^T \nonumber
\end{align}
The queries are generated at the user in the following way. First, the user generates $\frac{\bar{L}}{F}$ many independent and random  column vectors $U_1, U_2, \cdots, U_{\frac{\bar{L}}{F}}$, each with length $KL$. The queries generated by the users are given as
\begin{align}
\begin{bmatrix} \bar{Q}_1 & \bar{Q}_2 & \cdots & \bar{Q}_{\bar{L}}\end{bmatrix} \triangleq \begin{bmatrix} U_1 & U_2 & \cdots & U_{\frac{\bar{L}}{F}} \end{bmatrix} \cdot \mathbf{G}_{\left(\bar{L}, \frac{\bar{L}}{F} \right)}+ \begin{bmatrix} \mathbf{0} &\cdots & \mathbf{0} & e_1^{[\theta]} & \cdots & e_{L}^{[\theta]} \end{bmatrix} \label{Nan01}
\end{align}
where $\mathbf{G}_{\left(\bar{L}, \frac{\bar{L}}{F} \right)}$ is the generating matrix of an $\left(\bar{L}, \frac{\bar{L}}{F} \right)$-GRS code, $e_k^{[\theta]}$ is a column vector of length $KL$ where only the $\left((\theta-1)L+k\right)$-th element is $1$, and all other elements are $0$. The number of zero vectors in the second term of the RHS of (\ref{Nan01}) is $\bar{L}-L$. The total number of query column vectors in (\ref{Nan01}) is $\bar{L}$, and these will be distributed to the servers according to $(\mathbf{y}, F)$, more specifically, the number of queries to server $n$ is $\frac{y_n}{F} \bar{L}$, i.e., server $n$ receives $\bar{Q}_n$, $n \in \left[\sum_{i=1}^{n-1}\frac{y_i}{F} \bar{L}+1:\sum_{i=1}^{n}\frac{y_i}{F} \bar{L} \right]$ as query vectors.

We now show that the queries generated above protects the user's privacy, i.e., it satisfies (\ref{PrivacyConstraint}). The number of query vectors each colluding set $\mathcal{T}_m, m \in [1:M_c]$ sees is
\begin{align}
\sum_{n \in \mathcal{T}_m} \frac{y_n}{F} \bar{L} =\frac{\bar{L}}{F} \sum_{n \in \mathcal{T}_m} y_n \leq \frac{\bar{L}}{F}  \label{Nan03}
\end{align}
The last inequality of (\ref{Nan03}) is true because either $\mathcal{T}_m$ is in the joint pattern $\mathcal{P}_J$, or it is a subset of some eavesdropping set, which is in $\mathcal{P}_J$. Eitherway, since $\mathbf{y}$ satisfies (\ref{Constraint02}), we have $\sum_{n \in \mathcal{T}_m} y_n \leq 1$. Hence, we conclude that the number of query vectors each colluding set of server sees is no more than $\frac{\bar{L}}{F}$. Due to the MDS property of the $\left(\bar{L}, \frac{\bar{L}}{F} \right)$-GRS code, any $\frac{\bar{L}}{F}$ number of column vectors in the LHS of (\ref{Nan01}) are independent and uniformly distributed. Hence, user privacy is preserved against any colluding set of servers $\mathcal{T}_m, m \in [1:M_c]$. 

We note here that by a similar argument, the passive eavesdropper who observes an eavesdropping set $\mathcal{E}_m$, $m \in [1:M_e]$, from the eavesdropping pattern $\mathcal{P}_e$ does not know anything about the user's index of interest, and hence, the scheme preserves the user's privacy against the passive eavesdropper as well. 

Next, we describe server answers based on the query vector it receives. The $N$ servers share a common randomness vector $\mathbf{S}=\begin{bmatrix} S_1 & S_2 & \cdots &S_{\frac{\bar{L}}{F}}\end{bmatrix}$. Thus, according to (\ref{Nan05}), the amount of common randomness shared relative to the message size is 
\begin{align}
\rho=\frac{H(\mathbf{S})}{L}=\frac{\frac{\bar{L}}{F}}{\bar{L} \left(1-\frac{1}{F} \right)}=\frac{1}{F-1} \nonumber
\end{align}
 Each server calculates the following row vector
$
\bar{\mathbf{S}}=\begin{bmatrix} \bar{S}_1 & \bar{S}_2 & \cdots &\bar{S}_{\bar{L}}\end{bmatrix} \triangleq \mathbf{S} \cdot \mathbf{G}_{\left(\bar{L}, \frac{\bar{L}}{F} \right)}
$ of size $1 \times \bar{L}$. Server $n$ calculates the inner product of each query vector it receives with the message vector $\mathbf{W}$, and then adds the corresponding elements from $\bar{\mathbf{S}}$ and sends it to the user, i.e., let 
\begin{align}
\bar{A}_k \triangleq \bar{Q}_k^T \mathbf{W}+\bar{S}_k, \qquad k \in \left[1:\bar{L} \right], \label{NanNan02}
\end{align}
Server $n$ sends
$
\bar{A}_k, k \in \left[\sum_{i=1}^{n-1}\frac{y_i}{F} \bar{L}+1:\sum_{i=1}^{n}\frac{y_i}{F} \bar{L} \right] 
$ back to the user.

Now, we verify that the proposed scheme satisfies the decoding constraint at the user, i.e., (\ref{NoError}), the database privacy constraint at the user, i.e., (\ref{DatabasePrivacyUser}) and the database privacy constraint at the passive eavesdropper, i.e., (\ref{EC}). To do this, it is useful to write the server answers in a different form by noticing that both the random vectors $U_1, U_2, \cdots, U_{\frac{\bar{L}}{F}}$ and the common randomness $S_1, S_2, \cdots, S_{\frac{\bar{L}}{F}}$ are encoded by the same $\left(\bar{L}, \frac{\bar{L}}{F} \right)$-GRS code with generating matrix $\mathbf{G}_{\left(\bar{L}, \frac{\bar{L}}{F} \right)}$, i.e., the answers sent by the servers may be written in the following matrix form
\begin{align}
&\begin{bmatrix} \bar{A}_1 & \bar{A}_2 & \cdots & \bar{A}_{\bar{L}}\end{bmatrix} \nonumber\\
= & \begin{bmatrix} \mathbf{W}^T U_1+S_1 & \mathbf{W}^T U_2+S_2 & \cdots & \mathbf{W}^T U_{\frac{\bar{L}}{F}}+S_{\frac{\bar{L}}{F}} \end{bmatrix} \cdot \mathbf{G}_{\left(\bar{L}, \frac{\bar{L}}{F} \right)}+\begin{bmatrix} 0  & \cdots & 0 & W_\theta^1  & \cdots & W_\theta^L \end{bmatrix}
\nonumber \\
= & \begin{bmatrix} X_1 &\cdots &X_{\frac{\bar{L}}{F}} &W_\theta^1  & \cdots & W_\theta^L  \end{bmatrix} \cdot \begin{bmatrix} \mathbf{G}_{\left(\bar{L}, \frac{\bar{L}}{F} \right)}\\ \mathbf{0}_{L \times \frac{\bar{L}}{F}} \quad \mathbf{I}_L\end{bmatrix}\nonumber
\end{align}
where we have defined $X_l=\mathbf{W}^T U_l+S_l$, $ l \in \left[1:\frac{\bar{L}}{F} \right]$. 

The decoding constraint at the user is satisfied, as upon receiving $\bar{A}_l$, $l \in [1:\bar{L}]$, the user may decode $\begin{bmatrix} X_1 &\cdots &X_{\frac{\bar{L}}{F}} &W_\theta^1  & \cdots & W_\theta^L  \end{bmatrix}$ because $\begin{bmatrix} \mathbf{G}_{\left(\bar{L}, \frac{\bar{L}}{F} \right)}\\ \mathbf{0}_{L \times \frac{\bar{L}}{F}} \quad \mathbf{I}_L\end{bmatrix}$ is invertible due to the MDS property of the generating matrix $\mathbf{G}_{\left(\bar{L}, \frac{\bar{L}}{F} \right)}$. The database privacy against the user is satisfied because apart from its decoded message $W_\theta$, it can further decode $X_l$, $l \in \left[1:\frac{\bar{L}}{F} \right]$, which includes a common randomness $S_l$ to protect the other $K-1$ messages.  Finally, the proposed scheme protects database privacy against the passive eavesdropper because the number of answers each eavesdropping set $\mathcal{E}_m$, $m \in [1:M_e]$ sees is
\begin{align}
\sum_{n \in \mathcal{E}_m} \frac{y_n}{F} \bar{L} =\frac{\bar{L}}{F} \sum_{n \in \mathcal{E}_m} y_n \leq \frac{\bar{L}}{F}  \nonumber
\end{align}
which follows similar reasoning as (\ref{Nan03}). From (\ref{NanNan02}), we see that each answer $\bar{A}_k$ is protected by a randomness $\bar{S}_k$, and as long as the eavesdropper sees no more than $ \frac{\bar{L}}{F}$ many answers, the corresponding $ \frac{\bar{L}}{F}$ number of $\bar{S}$s are linearly independent, due to the MDS property of the generating matrix $\mathbf{G}_{\left(\bar{L}, \frac{\bar{L}}{F} \right)}$. Hence, the  inner product of the corresponding query vector and message vector is fully protected against the passive eavesdropper who observes answers from servers in $\mathcal{E}_m$.

\section{Proof of Theorem \ref{TheoremConverseAnyX}} \label{proof01}

 Define $\mathcal{Q}$ as the complete set of queries, i.e., $\mathcal{Q}=\{Q_n^{[k]}|n \in [1:N], k \in [1:K]\}$. 
 Using standard SPIR converse techniques in \cite{wang2019adversaries}, we can obtain 
\begin{align}
H(S) \geq H(A_\mathcal{N}^{[k]}|\mathcal{Q}) \label{idea01}
\end{align}
and
\begin{align}
H(W_k) \leq H(A_{[1:N]}^{[k]}|\mathcal{Q})-H(A_{\mathcal{N}}^{[k]}|\mathcal{Q}) \label{Nan08}
\end{align}
for any $\mathcal{N} \subset [1:N]$ that is a collusion set in $\mathcal{P}_c$ or an eavesdropping set in $\mathcal{P}_e$, i.e., in SPIR problems, the following equality holds true for $\mathcal{N}$ whether it is a colluding set or an eavedropping set  \cite{wang2019adversaries} :
\begin{align}
H(A_{\mathcal{N}}^{[k]}|W_k, Q_{\mathcal{N}}^{[k]})=H(A_\mathcal{N}^{[k]}|Q_{\mathcal{N}}^{[k]}). \label{AllSatisfy}
\end{align}
The fact that (\ref{AllSatisfy}) holds for all eavesdropping sets $\mathcal{N} \in \mathcal{P}_e$  is obvious, as for any passive eavesdropper, observing the queries to and answers from $\mathcal{N}$ servers should tell nothing about $W_k$, i.e., $\left(Q_{\mathcal{N}}^{[k]}, A_{\mathcal{N}}^{[k]} \right)$ is independent to $W_k$, $\forall k \in [1:K]$. As for a colluding set $\mathcal{N}  \in \mathcal{P}_c$, we have
\begin{align}
H(A_\mathcal{N}^{[k]}|W_k, Q_\mathcal{N}^{[k]})=H(A_\mathcal{N}^{[k']}|W_k, Q_\mathcal{N}^{[k']})=H(A_\mathcal{N}^{[k']}|Q_\mathcal{N}^{[k']})=H(A_\mathcal{N}^{[k]}|Q_\mathcal{N}^{[k]}) \nonumber
\end{align}
the first and third equality follows from protecting user's privacy against the server, i.e., (\ref{PrivacyConstraint}), while the second equality follows from protecting the database's privacy against the user, i.e., (\ref{DatabasePrivacyUser}). Hence, we conclude that (\ref{idea01}) and (\ref{Nan08}) both hold when $\mathcal{N} \in \mathcal{P}_J$. 

Similar to \cite{Liu2020ACPIR}, for each server set $\mathcal{N}_m \in \mathcal{P}_J$, $m \in [1:M_J]$, multiply both sides of (\ref{idea01}) by $x_m$, which is the $m$-th element of $\mathbf{x}$. Note that $\mathbf{x}$ satisfies (\ref{Positive02}), which means that we are multiplying non-negative numbers and the sign of the inequality does not need to be changed. Then, adding all these $M_J$ inequalities together, we obtain
\begin{align}
F_2 \cdot H(S) &\geq \sum_{m=1}^{M_J} x_m  H(A_{\mathcal{N}_{m}}^{[k]}|\mathcal{Q}), \nonumber
\end{align}
where we have used the definition of $F_2$, i.e., $F_2=\sum_{m=1}^M x_m$. Based on the results in \cite{Liu2020ACPIR}, we have 
\begin{align}
\sum_{m=1}^{M_J} x_m  H\left(A_{\mathcal{N}_{m}}^{[k]}|\mathcal{Q}\right) \geq H\left(A_{[1:N]}^{[k]}|\mathcal{Q}\right). \label{Nan09}
\end{align}
Hence, we have shown that 
\begin{align}
H(S) \geq \frac{1}{F_2} H\left(A_{[1:N]}^{[k]}|\mathcal{Q}\right) \label{Nan12}
\end{align}
which tells us that in order to protect the user's privacy against colluding server sets in $\mathcal{P}_c$, the database's privacy against the user, and the database's privacy against the passive eavesdropping on the sets in $\mathcal{P}_e$, the amount of common randomness must be no smaller than the ratio of the downloaded amount and $F_2$. 

Similarly, for each server set $\mathcal{N}_m \in \mathcal{P}_J$, $m \in [1:M_J]$, multiply both sides of (\ref{Nan08}) by $x_m$, which is the $m$-th element of $\mathbf{x}$, and adding all these $M_J$ inequalities together, we obtain
\begin{align}
F_2 \cdot H(W_k) &\leq F_2 \cdot H(A_{[1:N]}^{[k]}|\mathcal{Q})-\sum_{m=1}^{M_J} x_m H(A_{\mathcal{N}}^{[k]}|\mathcal{Q}) \nonumber\\
& \leq F_2 \cdot H(A_{[1:N]}^{[k]}|\mathcal{Q})-H\left(A_{[1:N]}^{[k]}|\mathcal{Q}\right) \label{Nan10}
\end{align}
where (\ref{Nan10}) follows from (\ref{Nan09}). Thus, we have
\begin{align}
H(A_{1:N}^{[k]}|\mathcal{Q}) \geq \frac{F_2}{F_2-1} H(W_k) \label{Nan13}
\end{align}
which combined with (\ref{Nan12}) gives us a lower bound on the common randomness relative to the message size, i.e.,
\begin{align}
\rho=\frac{H(S)}{H(W_k)} \geq \frac{1}{F_2-1}. 
\end{align}

Using (\ref{Nan13}), we obtain an upper bound on the capacity of the SPIR problem under collusion pattern $\mathcal{P}_c$ and $\mathcal{P}_e$, i.e., 
\begin{align}
C(\mathcal{P}_c,\mathcal{P}_e) =\frac{H(W_k)}{\sum_{n=1}^N H(A_n^{[k]} )} \leq \frac{H(W_k)}{\sum_{n=1}^N H(A_n^{[k]}|\mathcal{Q} )} \leq \frac{H(W_k)}{ H(A_{1:N}^{[k]}|\mathcal{Q} )} \leq  1-\frac{1}{F_2}.
\end{align}

\bibliographystyle{unsrt}
\bibliography{main}

\begin{thebibliography}{100}

\bibitem{chor1995private}
Benny Chor, Oded Goldreich, Eyal Kushilevitz, and Madhu Sudan.
\newblock Private information retrieval.
\newblock In {\em Proceedings of IEEE 36th Annual Foundations of Computer
  Science}, pages 41--50, Oct. 1995.

\bibitem{sun2017capacity}
Hua Sun and Syed~Ali Jafar.
\newblock The capacity of private information retrieval.
\newblock {\em IEEE Transactions on Information Theory}, 63(7):4075--4088, Jul.
  2017.

\bibitem{2000Protecting}
Yael Gertner, Yuval Ishai, and Eyal Kushilevitz.
\newblock Protecting data privacy in private information retrieval schemes.
\newblock {\em Journal of computer and system ences}, 60(3):p.592--629, 2000.

\bibitem{Sun2017SPIR}
H.~{Sun} and S.~A. {Jafar}.
\newblock The capacity of symmetric private information retrieval.
\newblock {\em IEEE Transactions on Information Theory}, 65(1):322--329, 2019.

\bibitem{sun2018colluding}
Hua Sun and Syed~Ali Jafar.
\newblock The capacity of robust private information retrieval with colluding
  databases.
\newblock {\em IEEE Transactions on Information Theory}, 64(4):2361--2370, Apr.
  2018.

\bibitem{wang2018secure}
Qiwen Wang and Mikael Skoglund.
\newblock Secure private information retrieval from colluding databases with
  eavesdroppers.
\newblock In {\em 2018 IEEE International Symposium on Information Theory
  (ISIT)}, pages 2456--2460, Jun. 2018.

\bibitem{wang2019adversaries}
Qiwen Wang and Mikael Skoglund.
\newblock On {PIR} and symmetric {PIR} from colluding databases with
  adversaries and eavesdroppers.
\newblock {\em IEEE Transactions on Information Theory}, 65(5):3183--3197, May
  2019.

\bibitem{Wang2019MDSTSPIR}
Q.~{Wang} and M.~{Skoglund}.
\newblock Symmetric private information retrieval from mds coded distributed
  storage with non-colluding and colluding servers.
\newblock {\em IEEE Transactions on Information Theory}, 65(8):5160--5175,
  2019.

\bibitem{zhang2019private}
Yiwei Zhang, Xin Wang, Hengjia Wei, and Gennian Ge.
\newblock On private information retrieval array codes.
\newblock {\em IEEE Transactions on Information Theory}, 65(9):5565--5573, Sep.
  2019.

\bibitem{tandon2017capacity}
Ravi Tandon.
\newblock The capacity of cache aided private information retrieval.
\newblock In {\em 2017 55th Annual Allerton Conference on Communication,
  Control, and Computing (Allerton)}, pages 1078--1082, Oct. 2017.

\bibitem{banawan2018capacity2}
Karim Banawan and Sennur Ulukus.
\newblock The capacity of private information retrieval from byzantine and
  colluding databases.
\newblock {\em IEEE Transactions on Information Theory}, 65(2):1206--1219, Feb.
  2019.

\bibitem{tajeddine2017robust}
Razane Tajeddine and Salim~El Rouayheb.
\newblock Robust private information retrieval on coded data.
\newblock In {\em 2017 IEEE International Symposium on Information Theory
  (ISIT)}, pages 1903--1907, Jun. 2017.

\bibitem{bitar2018staircase}
Rawad Bitar and Salim~El Rouayheb.
\newblock Staircase-{PIR}: Universally robust private information retrieval.
\newblock In {\em 2018 IEEE Information Theory Workshop (ITW)}, Nov. 2018.

\bibitem{fanti2015efficient}
Giulia Fanti and Kannan Ramchandran.
\newblock Efficient private information retrieval over unsynchronized
  databases.
\newblock {\em IEEE Journal of Selected Topics in Signal Processing},
  9(7):1229--1239, Oct. 2015.

\bibitem{wang2018capacity}
Qiwen Wang, Hua Sun, and Mikael Skoglund.
\newblock The capacity of private information retrieval with eavesdroppers.
\newblock {\em IEEE Transactions on Information Theory}, 65(5):3198--3214, May
  2018.

\bibitem{wang2019mismatched}
Qiwen Wang, Hua Sun, and Mikael Skoglund.
\newblock Symmetric private information retrieval with mismatched coded
  messages and randomness.
\newblock In {\em 2019 IEEE International Symposium on Information Theory
  (ISIT)}, Jul. 2019.

\bibitem{sun2018multiround}
Hua Sun and Syed~Ali Jafar.
\newblock Multiround private information retrieval: Capacity and storage
  overhead.
\newblock {\em IEEE Transactions on Information Theory}, 64(8):5743--5754, Aug.
  2018.

\bibitem{Xu2018Subpacket}
J.~Xu and Z.~Zhang.
\newblock On sub-packetization and access number of capacity-achieving {PIR}
  schemes for {MDS} coded non-colluding databases.
\newblock {\em SCIENCE CHINA Information Sciences}, 61(10):100 306:1–100
  306:16, Aug. 2018.

\bibitem{wei2019capacity}
Yi-Peng Wei, Karim Banawan, and Sennur Ulukus.
\newblock The capacity of private information retrieval with partially known
  private side information.
\newblock {\em IEEE Transactions on Information Theory}, Mar. 2019.

\bibitem{samy2019on}
Islam Samy, Ravi Tandon, and Loukas Lazos.
\newblock On the capacity of leaky private information retrieval.
\newblock In {\em 2019 IEEE International Symposium on Information Theory
  (ISIT)}, Jul. 2019.

\bibitem{zhou2020capacity-achieving}
Ruida Zhou, Chao Tian, Hua Sun, and Tie Liu.
\newblock Capacity-achieving private information retrieval codes from mds-coded
  databases with minimum message size.
\newblock {\em IEEE Transactions on Information Theory}, pages 1--1, 2020.

\bibitem{chen2020the}
Zhen Chen, Zhiying Wang, and Syed~A Jafar.
\newblock The capacity of t-private information retrieval with private side
  information.
\newblock {\em IEEE Transactions on Information Theory}, pages 1--1, 2020.

\bibitem{chen2020the2}
Zhen Chen, Zhiying Wang, and Syed~A Jafar.
\newblock The asymptotic capacity of private search.
\newblock {\em IEEE Transactions on Information Theory}, pages 1--1, 2020.

\bibitem{chen2020gcsa}
Zhen Chen, Zhuqing Jia, Zhiying Wang, and Syed~A Jafar.
\newblock Gcsa codes with noise alignment for secure coded multi-party batch
  matrix multiplication.
\newblock {\em arXiv: Information Theory}, 2020.

\bibitem{xu2019capacity}
Jingke Xu, Yaqian Zhang, and Zhifang Zhang.
\newblock A capacity-achieving {$ T $-PIR} scheme based on {MDS} array codes.
\newblock {\em arXiv preprint arXiv:1901.05772}, Jan. 2019.

\bibitem{wei2018capacity}
Yi-Peng Wei, Batuhan Arasli, Karim Banawan, and Sennur Ulukus.
\newblock The capacity of private information retrieval from decentralized
  uncoded caching databases.
\newblock {\em arXiv preprint arXiv:1811.11160}, Nov. 2018.

\bibitem{chee2019generalization}
Yeow~Meng Chee, Han~Mao Kiah, Eitan Yaakobi, and Hui Zhang.
\newblock A generalization of the blackburn-etzion construction for private
  information retrieval array codes.
\newblock In {\em 2019 IEEE International Symposium on Information Theory
  (ISIT)}, Jul. 2019.

\bibitem{lin2019improved}
Hsuan-Yin Lin, Siddhartha Kumar, and Eirik Rosnes.
\newblock Improved private information retrieval for coded storage from code
  decomposition.
\newblock In {\em 2019 IEEE Information Theory Workshop (ITW)}. IEEE, Aug.
  2019.

\bibitem{xu2018building}
Jingke Xu and Zhifang Zhang.
\newblock Building capacity-achieving {PIR} schemes with optimal
  sub-packetization over small fields.
\newblock In {\em 2018 IEEE International Symposium on Information Theory
  (ISIT)}, pages 1749--1753. IEEE, Jun. 2018.

\bibitem{tian2018shannon}
Chao Tian, Hua Sun, and Jun Chen.
\newblock A shannon-theoretic approach to the storage-retrieval tradeoff in
  {PIR} systems.
\newblock In {\em 2018 IEEE International Symposium on Information Theory
  (ISIT)}, pages 1904--1908. IEEE, Jun. 2018.

\bibitem{fazeli2015codes}
Arman Fazeli, Alexander Vardy, and Eitan Yaakobi.
\newblock Codes for distributed {PIR} with low storage overhead.
\newblock In {\em 2015 IEEE International Symposium on Information Theory
  (ISIT)}, pages 2852--2856. IEEE, Jun. 2015.

\bibitem{blackburn2019pir}
Simon~R Blackburn and Tuvi Etzion.
\newblock {PIR} array codes with optimal virtual server rate.
\newblock {\em IEEE Transactions on Information Theory}, 65(10):6136--6145,
  Oct. 2019.

\bibitem{blackburn2017pir2}
Simon~R Blackburn, Tuvi Etzion, and Maura~B Paterson.
\newblock {PIR} schemes with small download complexity and low storage
  requirements.
\newblock In {\em 2017 IEEE International Symposium on Information Theory
  (ISIT)}, pages 146--150, Jun. 2017.

\bibitem{kumar2018local}
Siddhartha Kumar, Hsuan-Yin Lin, Eirik Rosnes, and Alexandre Graell~iAmat.
\newblock Local reconstruction codes: A class of {MDS-PIR} capacity-achieving
  codes.
\newblock In {\em 2018 IEEE Information Theory Workshop (ITW)}. IEEE, Nov.
  2018.

\bibitem{blackburn2017pir}
Simon~R Blackburn and Tuvi Etzion.
\newblock {PIR} array codes with optimal {PIR} rates.
\newblock In {\em 2017 IEEE International Symposium on Information Theory
  (ISIT)}, pages 2658--2662. IEEE, Jun. 2017.

\bibitem{lavauzelle2018private}
Julien Lavauzelle.
\newblock Private information retrieval from transversal designs.
\newblock {\em IEEE Transactions on Information Theory}, 65(2):1189--1205, Feb.
  2018.

\bibitem{chan2015private}
Terence~H Chan, Siu-Wai Ho, and Hirosuke Yamamoto.
\newblock Private information retrieval for coded storage.
\newblock In {\em 2015 IEEE International Symposium on Information Theory
  (ISIT)}, pages 2842--2846. IEEE, Jun. 2015.

\bibitem{yang2002private}
Erica~Y Yang, Jie Xu, and Keith~H Bennett.
\newblock Private information retrieval in the presence of malicious failures.
\newblock In {\em Proceedings 26th Annual International Computer Software and
  Applications}, pages 805--810. IEEE, Aug. 2002.

\bibitem{sun2016blind}
Hua Sun and Syed~A Jafar.
\newblock Blind interference alignment for private information retrieval.
\newblock In {\em 2016 IEEE International Symposium on Information Theory
  (ISIT)}, pages 560--564. IEEE, Jul. 2016.

\bibitem{melchor2008fast}
Carlos~Aguilar Melchor and Philippe Gaborit.
\newblock A fast private information retrieval protocol.
\newblock In {\em 2008 IEEE International Symposium on Information Theory},
  pages 1848--1852. IEEE, Jul. 2008.

\bibitem{fanti2014multi}
Giulia Fanti and Kannan Ramchandran.
\newblock Multi-server private information retrieval over unsynchronized
  databases.
\newblock In {\em 2014 52nd Annual Allerton Conference on Communication,
  Control, and Computing (Allerton)}, pages 437--444. IEEE, Sep. 2014.

\bibitem{shah2014one}
Nihar~B Shah, KV~Rashmi, and Kannan Ramchandran.
\newblock One extra bit of download ensures perfectly private information
  retrieval.
\newblock In {\em 2014 IEEE International Symposium on Information Theory},
  pages 856--860. IEEE, Jun. 2014.

\bibitem{banawan2018private2}
Karim Banawan and Sennur Ulukus.
\newblock Private information retrieval from multiple access channels.
\newblock In {\em 2018 IEEE Information Theory Workshop (ITW)}, Nov. 2018.

\bibitem{jia2019cross}
Zhuqing Jia, Hua Sun, and Syed~A Jafar.
\newblock Cross subspace alignment and the asymptotic capacity of {$X$}-secure
  {$T$}-private information retrieval.
\newblock {\em IEEE Transactions on Information Theory}, 65(9):5783--5798, Sep.
  2019.

\bibitem{kumar2019achieving}
Siddhartha Kumar, Hsuan-Yin Lin, Eirik Rosnes, and Alexandre~Graell i~Amat.
\newblock Achieving maximum distance separable private information retrieval
  capacity with linear codes.
\newblock {\em IEEE Transactions on Information Theory}, 65(7):4243--4273, Jul.
  2019.

\bibitem{vajha2017binary}
Myna Vajha, Vinayak Ramkumar, and P~Vijay Kumar.
\newblock Binary, shortened projective reed muller codes for coded private
  information retrieval.
\newblock In {\em 2017 IEEE International Symposium on Information Theory
  (ISIT)}, pages 2648--2652. IEEE, Jun. 2017.

\bibitem{kadhe2017private}
Swanand Kadhe, Brenden Garcia, Anoosheh Heidarzadeh, Salim El~Rouayheb, and
  Alex Sprintson.
\newblock Private information retrieval with side information: The single
  server case.
\newblock In {\em 2017 55th Annual Allerton Conference on Communication,
  Control, and Computing (Allerton)}, pages 1099--1106. IEEE, Oct. 2017.

\bibitem{wei2018private}
Yi-Peng Wei and Sennur Ulukus.
\newblock Private information retrieval with private side information under
  storage constraints.
\newblock In {\em 2018 IEEE Information Theory Workshop (ITW)}. IEEE, Nov.
  2018.

\bibitem{li2018single}
Su~Li and Michael Gastpar.
\newblock Single-server multi-user private information retrieval with side
  information.
\newblock In {\em 2018 IEEE International Symposium on Information Theory
  (ISIT)}, pages 1954--1958. IEEE, Jun. 2018.

\bibitem{schrijver2003combinatorial}
Alexander Schrijver.
\newblock {\em Combinatorial optimization: polyhedra and efficiency},
  volume~24.
\newblock Springer Science \& Business Media, 2003.

\bibitem{banawan2019improved}
Karim Banawan, Batuhan Arasli, and Sennur Ulukus.
\newblock Improved storage for efficient private information retrieval.
\newblock {\em arXiv preprint arXiv:1908.11366}, Aug. 2019.

\bibitem{jia2019x}
Zhuqing Jia and Syed~A Jafar.
\newblock {$X$}-secure {$T$}-private information retrieval from {MDS} coded
  storage with byzantine and unresponsive servers.
\newblock {\em arXiv preprint arXiv:1908.10854}, Aug. 2019.

\bibitem{sun2019breaking}
Hua Sun and Chao Tian.
\newblock Breaking the {MDS-PIR} capacity barrier via joint storage coding.
\newblock {\em Information}, 10(9):265, 2019.

\bibitem{kadhe2019equivalence}
Swanand Kadhe, Anoosheh Heidarzadeh, Alex Sprintson, and O~Ozan Koyluoglu.
\newblock On an equivalence between single-server {PIR} with side information
  and locally recoverable codes.
\newblock {\em arXiv preprint arXiv:1907.00598}, Jul. 2019.

\bibitem{kazemi2019private}
Fatemeh Kazemi, Esmaeil Karimi, Anoosheh Heidarzadeh, and Alex Sprintson.
\newblock Private information retrieval with private coded side information:
  The multi-server case.
\newblock {\em arXiv preprint arXiv:1906.11278}, Jun. 2019.

\bibitem{jia2019asymptotic}
Zhuqing Jia and Syed~A Jafar.
\newblock On the asymptotic capacity of {$X$}-secure {$T$}-private information
  retrieval with graph based replicated storage.
\newblock {\em arXiv preprint arXiv:1904.05906}, Apr. 2019.

\bibitem{woolsey2019optimal}
Nicholas Woolsey, Rong-Rong Chen, and Mingyue Ji.
\newblock An optimal iterative placement algorithm for {PIR} from heterogeneous
  storage-constrained databases.
\newblock {\em arXiv preprint arXiv:1904.02131}, Apr. 2019.

\bibitem{zhou2019capacity}
Ruida Zhou, Chao Tian, Hua Sun, and Tie Liu.
\newblock Capacity-achieving private information retrieval codes from
  {MDS}-coded databases with minimum message size.
\newblock {\em arXiv preprint arXiv:1903.08229}, Mar. 2019.

\bibitem{zhu2019new}
Jinbao Zhu, Qifa Yan, Chao Qi, and Xiaohu Tang.
\newblock A new capacity-achieving private information retrieval scheme with
  (almost) optimal file length for coded servers.
\newblock {\em IEEE Transactions on Information Forensics and Security}, Aug.
  2019.

\bibitem{banawan2019capacity}
Karim Banawan, Batuhan Arasli, Yi-Peng Wei, and Sennur Ulukus.
\newblock The capacity of private information retrieval from heterogeneous
  uncoded caching databases.
\newblock {\em arXiv preprint arXiv:1902.09512}, Feb. 2019.

\bibitem{heidarzadeh2018capacityITW}
Anoosheh Heidarzadeh, Fatemeh Kazemi, and Alex Sprintson.
\newblock Capacity of single-server single-message private information
  retrieval with coded side information.
\newblock In {\em 2018 IEEE Information Theory Workshop (ITW)}, 2018.

\bibitem{kazemi2019single}
Fatemeh Kazemi, Esmaeil Karimi, Anoosheh Heidarzadeh, and Alex Sprintson.
\newblock Single-server single-message online private information retrieval
  with side information.
\newblock {\em arXiv preprint arXiv:1901.07748}, Jan. 2019.

\bibitem{heidarzadeh2019single}
Anoosheh Heidarzadeh, Swanand Kadhe, Salim~El Rouayheb, and Alex Sprintson.
\newblock Single-server multi-message individually-private information
  retrieval with side information.
\newblock {\em arXiv preprint arXiv:1901.07509}, Feb. 2019.

\bibitem{woolsey2019new}
Nicholas Woolsey, Rong-Rong Chen, and Mingyue Ji.
\newblock A new design of private information retrieval for storage constrained
  databases.
\newblock {\em arXiv preprint arXiv:1901.07490}, Jan. 2019.

\bibitem{Hsuan2019Weakly}
Hsuan-Yin Lin, Siddhartha Kumar, and Eirik Rosnes.
\newblock Weakly-private information retrieval.
\newblock {\em arXiv preprint arXiv:1901.06730}, May. 2019.

\bibitem{OurArXiv}
Xinyu Yao, Nan Liu, and Wei Kang.
\newblock The capacity of multi-round private information retrieval from
  byzantine databases.
\newblock In {\em 2019 IEEE International Symposium on Information Theory
  (ISIT)}, Jul. 2019.

\bibitem{macwilliams1977theory}
Florence~Jessie MacWilliams and Neil James~Alexander Sloane.
\newblock {\em The theory of error-correcting codes}.
\newblock Elsevier, 1977.

\bibitem{feyling1993punctured}
C~Feyling.
\newblock Punctured maximum distance separable codes.
\newblock {\em Electronics Letters}, 29(5):470--471, Mar. 1993.

\bibitem{ling2004coding}
San Ling and Chaoping Xing.
\newblock {\em Coding theory: a first course}.
\newblock Cambridge University Press, 2004.

\bibitem{van2012introduction}
Jacobus~Hendricus Van~Lint.
\newblock {\em Introduction to coding theory}, volume~86.
\newblock Springer Science \& Business Media, 2012.

\bibitem{wang2018epsilon}
Qiwen Wang, Hua Sun, and Mikael Skoglund.
\newblock The $\epsilon $-error capacity of symmetric {PIR} with byzantine
  adversaries.
\newblock In {\em 2018 IEEE Information Theory Workshop (ITW)}, Sep. 2018.

\bibitem{wang2017secure}
Qiwen Wang and Mikael Skoglund.
\newblock Secure symmetric private information retrieval from colluding
  databases with adversaries.
\newblock In {\em 2017 55th Annual Allerton Conference on Communication,
  Control, and Computing (Allerton)}, pages 1083--1090, Oct. 2017.

\bibitem{tajeddine2018robust}
Razane Tajeddine, Oliver~W Gnilke, David Karpuk, Ragnar Freij-Hollanti, and
  Camilla Hollanti.
\newblock Robust private information retrieval from coded systems with
  byzantine and colluding servers.
\newblock In {\em 2018 IEEE International Symposium on Information Theory
  (ISIT)}, pages 2451--2455, Jun. 2018.

\bibitem{raviv2018private}
Netanel Raviv and Itzhak Tamot.
\newblock Private information retrieval in graph based replication systems.
\newblock In {\em 2018 IEEE International Symposium on Information Theory
  (ISIT)}, pages 1739--1743, Jun. 2018.

\bibitem{kumar2018private}
Siddhartha Kumar, Alexandre Graell~i Amat, Eirik Rosnes, and Linda
  Senigagliesi.
\newblock Private information retrieval from a cellular network with caching at
  the edge.
\newblock {\em IEEE Transactions on Communications}, 67(7):4900--4912, July
  2019.

\bibitem{yang2018private}
Heecheol Yang, Wonjae Shin, and Jungwoo Lee.
\newblock Private information retrieval for secure distributed storage systems.
\newblock {\em IEEE Transactions on Information Forensics and Security},
  13(12):2953--2964, Dec. 2018.

\bibitem{kim2017cache}
Minchul Kim, Heecheol Yang, and Jungwoo Lee.
\newblock Cache-aided private information retrieval.
\newblock In {\em 2017 51st Asilomar Conference on Signals, Systems, and
  Computers}, pages 398--402, Oct. 2017.

\bibitem{sun2017optimal}
Hua Sun and Syed~Ali Jafar.
\newblock Optimal download cost of private information retrieval for arbitrary
  message length.
\newblock {\em IEEE Transactions on Information Forensics and Security},
  12(12):2920--2932, Dec. 2017.

\bibitem{d2018lifting}
Rafael~GL D'Oliveira and Salim~El Rouayheb.
\newblock Lifting private information retrieval from two to any number of
  messages.
\newblock In {\em 2018 IEEE International Symposium on Information Theory
  (ISIT)}, pages 1744--1748, Jun. 2018.

\bibitem{tajeddine2018private2}
Razane Tajeddine, Antonia Wachter-Zeh, and Camilla Hollanti.
\newblock Private information retrieval over random linear networks.
\newblock {\em IEEE Transactions on Information Forensics and Security}, pages
  790--799, Jan. 2020.

\bibitem{tian2018capacity}
Chao Tian, Hua Sun, and Jun Chen.
\newblock Capacity-achieving private information retrieval codes with optimal
  message size and upload cost.
\newblock In {\em ICC 2019 - 2019 IEEE International Conference on
  Communications (ICC)}, May. 2019.

\bibitem{banawan2018private}
Karim Banawan and Sennur Ulukus.
\newblock Private information retrieval through wiretap channel {II}.
\newblock In {\em 2018 IEEE International Symposium on Information Theory
  (ISIT)}, pages 971--975, Jun. 2018.

\bibitem{wang2017linear}
Qiwen Wang and Mikael Skoglund.
\newblock Linear symmetric private information retrieval for {MDS} coded
  distributed storage with colluding servers.
\newblock In {\em 2017 IEEE Information Theory Workshop (ITW)}, pages 71--75,
  Nov. 2017.

\bibitem{banawan2018noisy}
Karim Banawan and Sennur Ulukus.
\newblock Noisy private information retrieval: On separability of channel
  coding and information retrieval.
\newblock {\em IEEE Transactions on Information Theory}, Jul. 2018.

\bibitem{lin2018asymmetry}
Hsuan-Yin Lin, Siddhartha Kumar, Eirik Rosnes, and Alexandre Graell~i Amat.
\newblock Asymmetry helps: Improved private information retrieval protocols for
  distributed storage.
\newblock In {\em 2018 IEEE Information Theory Workshop (ITW)}, Nov. 2018.

\bibitem{shariatpanahi2018multi}
Seyed~Pooya Shariatpanahi, Mahdi~Jafari Siavoshani, and Mohammad~Ali
  Maddah-Ali.
\newblock Multi-message private information retrieval with private side
  information.
\newblock {\em arXiv preprint arXiv:1805.11892}, May 2018.

\bibitem{banawan2018multi}
Karim Banawan and Sennur Ulukus.
\newblock Multi-message private information retrieval: Capacity results and
  near-optimal schemes.
\newblock {\em IEEE Transactions on Information Theory}, 64(10):6842--6862,
  Oct. 2018.

\bibitem{abdul2017private}
Ravi Tandon, Maryam Abdul-Wahid, Firas Almoualem, and Deepak Kumar.
\newblock Private information retrieval from storage constrained
  databases--coded caching meets {PIR}.
\newblock {\em 2018 IEEE International Conference on Communications (ICC)},
  2018.

\bibitem{heidarzadeh2018capacity}
Anoosheh Heidarzadeh, Brenden Garcia, Swanand Kadhe, Salim~El Rouayheb, and
  Alex Sprintson.
\newblock On the capacity of single-server multi-message private information
  retrieval with side information.
\newblock {\em arXiv preprint arXiv:1807.09908}, Jul. 2018.

\bibitem{chen2017capacity}
Zhen Chen, Zhiying Wang, and Syed Jafar.
\newblock The capacity of private information retrieval with private side
  information.
\newblock {\em arXiv preprint arXiv:1709.03022}, Sep. 2017.

\bibitem{wei2018cache}
Yi-Peng Wei, Karim Banawan, and Sennur Ulukus.
\newblock Cache-aided private information retrieval with partially known
  uncoded prefetching: Fundamental limits.
\newblock {\em IEEE Journal on Selected Areas in Communications},
  36(6):1126--1139, Jun. 2018.

\bibitem{wei2018fundamental}
Yi-Peng Wei, Karim Banawan, and Sennur Ulukus.
\newblock Fundamental limits of cache-aided private information retrieval with
  unknown and uncoded prefetching.
\newblock {\em IEEE Transactions on Information Theory}, 65(5):3215--3232, Feb.
  2018.

\bibitem{lin2018mds}
Hsuan-Yin Lin, Siddhartha Kumar, Eirik Rosnes, and Alexandre~Graell i~Amat.
\newblock An {MDS-PIR} capacity-achieving protocol for distributed storage
  using non-{MDS} linear codes.
\newblock In {\em 2018 IEEE International Symposium on Information Theory
  (ISIT)}, pages 966--970, Jun. 2018.

\bibitem{kumar2017private}
Siddhartha Kumar, Eirik Rosnes, and Alexandre~Graell i~Amat.
\newblock Private information retrieval in distributed storage systems using an
  arbitrary linear code.
\newblock In {\em 2017 IEEE International Symposium on Information Theory
  (ISIT)}, pages 1421--1425, Jun. 2017.

\bibitem{tajeddine2018private}
Razane Tajeddine, Oliver~W Gnilke, and Salim El~Rouayheb.
\newblock Private information retrieval from {MDS} coded data in distributed
  storage systems.
\newblock {\em IEEE Transactions on Information Theory}, 64(11):7081--7093,
  Nov. 2018.

\bibitem{Penas2019local}
Umberto Martinez-Penas.
\newblock Private information retrieval from locally repairable databases with
  colluding servers.
\newblock {\em arXiv preprint https://arxiv.org/pdf/1901.02938.pdf}, Jan. 2019.

\bibitem{Guo2019leak}
Tao Guo, Ruida Zhou, and Chao Tian.
\newblock On the information leakage in private information retrieval systems.
\newblock {\em arXiv preprint arXiv:1909.11605}, Sep. 2019.

\bibitem{Fazeli2015coded}
Arman~Fazeli andAlexander Vardy and Eitan Yaakobi.
\newblock Pir with low storage overhead: coding instead of replication.
\newblock {\em arXiv preprint https://arxiv.org/pdf/1505.06241.pdf}, May 2015.

\bibitem{Li2019repair}
Jie Li, David Karpuk, and Camilla Hollanti.
\newblock Private information retrieval from {MDS} array codes with (near-)
  optimal repair bandwidth.
\newblock {\em arXiv preprint arXiv:1909.10289}, Sep. 2019.

\bibitem{Holzbaur2019linear}
Lukas Holzbaur, Ragnar Freij-Hollanti, Jie Li, and Camilla Hollanti.
\newblock Capacity of linear private information retrieval from coded,
  colluding, and adversarial servers.
\newblock {\em arXiv preprint https://arxiv.org/pdf/1903.12552.pdf}, Mar. 2019.

\bibitem{zhang2017general}
Yiwei Zhang and Gennian Ge.
\newblock A general private information retrieval scheme for {MDS} coded
  databases with colluding servers.
\newblock {\em Designs, Codes and Cryptography}, May. 2019.

\bibitem{zhang2019subpacket}
Zhifang Zhang and Jingke Xu.
\newblock The optimal sub-packetization of linear capacity-achieving {PIR}
  schemes with colluding servers.
\newblock {\em IEEE Transactions on Information Theory}, 65(5):2723--2735, May
  2019.

\bibitem{tajeddine2019colluding}
Ragnar Freij-Hollanti, Oliver~W. Gnilke, Camilla Hollanti, Anna-Lena
  Horlemann-Trautmann, David Karpuk, and Ivo Kubjas.
\newblock {$T$}-private information retrieval schemes using transitive codes.
\newblock {\em IEEE Transactions on Information Theory}, 65(4):2107--2118, Apr.
  2019.

\bibitem{tajeddine2019byzantine}
Razane Tajeddine, Oliver~W. Gnilke, David Karpuk, Ragnar Freij-Hollanti, and
  Camilla Hollanti.
\newblock Private information retrieval from coded storage systems with
  colluding, byzantine, and unresponsive servers.
\newblock {\em IEEE Transactions on Information Theory}, 65(6):3898--3906, Jun.
  2019.

\bibitem{wang2019symmetric}
Qiwen Wang and Mikael Skoglund.
\newblock Symmetric private information retrieval from {MDS} coded distributed
  storage with non-colluding and colluding servers.
\newblock {\em IEEE Transactions on Information Theory}, 65(8):5160--5175, Aug.
  2019.

\bibitem{BAN18c}
Karim Banawan and Sennur Ulukus.
\newblock Private information retrieval from non-replicated databases.
\newblock {\em arXiv preprint arXiv:1901.00004}, Dec. 2018.

\bibitem{BAN18b}
Karim Banawan and Sennur Ulukus.
\newblock The capacity of private information retrieval from coded databases.
\newblock {\em IEEE Transactions on Information Theory}, 64(3):1945--1956, Mar.
  2018.

\bibitem{BAN18}
Karim Banawan and Sennur Ulukus.
\newblock Asymmetry hurts: Private information retrieval under asymmetric
  traffic constraints.
\newblock {\em IEEE Transactions on Information Theory}, pages 2446--2450, Feb.
  2018.

\bibitem{ATT18}
Mohamed~Adel Attia, Deepak Kumar, and Ravi Tandon.
\newblock The capacity of uncoded storage constrained {PIR}.
\newblock In {\em 2018 IEEE International Symposium on Information Theory
  (ISIT)}, pages 1959--1963. IEEE, Jun. 2018.

\bibitem{tajeddine2017arbitrary}
Razane Tajeddine, Oliver~W Gnilke, David Karpuk, Ragnar Freij-Hollanti, Camilla
  Hollanti, and Salim El~Rouayheb.
\newblock Private information retrieval schemes for coded data with arbitrary
  collusion patterns.
\newblock In {\em 2017 IEEE International Symposium on Information Theory
  (ISIT)}, pages 1908--1912, Jun. 2017.

\bibitem{zhang2017private}
Yiwei Zhang and Gennian Ge.
\newblock Private information retrieval from {MDS} coded databases with
  colluding servers under several variant models.
\newblock {\em arXiv preprint arXiv:1705.03186}, Oct. 2017.

\bibitem{jia2017disjoint}
Zhuqing Jia, Hua Sun, and Syed~A. Jafar.
\newblock The capacity of private information retrieval with disjoint colluding
  sets.
\newblock In {\em 2017 IEEE Global Communications Conference}, Dec. 2017.

\bibitem{sun2018private}
Hua Sun and Syed~Ali Jafar.
\newblock Private information retrieval from {MDS} coded data with colluding
  servers: Settling a conjecture by {F}reij-{H}ollanti et al.
\newblock {\em IEEE Transactions on Information Theory}, 64(2):1000--1022, Feb.
  2018.

\bibitem{Liu2020ACPIR}
N.~{Liu} X.~{Yao} and W.~{Kang}.
\newblock The capacity of private information retrieval under arbitrary
  collusion patterns.
\newblock {\em arXiv preprint arXiv}, (2001.03843), 2020.

\end{thebibliography}

\end{document}